    \documentclass[%
 reprint,
 superscriptaddress,
 amsmath,amssymb,
 aps,
prb,
floatfix,
longbibliography 
]{revtex4-2}

\usepackage{epsfig}
\usepackage{amssymb}
\usepackage{amsmath}
\usepackage{amsfonts}
\usepackage{braket}
\usepackage[T1]{fontenc}
\usepackage{bm}
\usepackage{graphicx}
\usepackage{xcolor}
\usepackage{verbatim}
\usepackage{soul}
\usepackage{tikz}
\usetikzlibrary{matrix,shapes,arrows,positioning,chains}
\usetikzlibrary { decorations.pathmorphing, decorations.pathreplacing, decorations.shapes,}
\usetikzlibrary{shapes.geometric}
\usetikzlibrary{shapes.arrows}
\usepackage{array}
\usepackage{enumitem}
\usepackage{color}
\usepackage{physics}
\usepackage[export]{adjustbox}
\usepackage[percent]{overpic}
\usepackage[normalem]{ulem}

\usepackage{bbold}

\usepackage{float}
\usepackage{dcolumn}
\usepackage{bm}
\usepackage[pdftex,colorlinks=true]{hyperref}
\hypersetup{
  colorlinks=true,
  linkcolor=blue,
  urlcolor=cyan,
}
\usepackage{tikz}
\usetikzlibrary{decorations.pathreplacing}

\newcommand{\stkout}[1]{\ifmmode\text{\sout{\ensuremath{#1}}}\else\sout{#1}\fi}

\newcommand{\eqa}[1]{\begin{align} #1 \end{align}}
\newcommand{\nn}{\nonumber}

\newcommand{\mD}{\mathcal{D}}

\newcommand{\pd}{\partial}
\newcommand{\bS}{\boldsymbol{S}}

\newcommand{\bx}{\boldsymbol{\hat{x}}}

\newcommand{\hS}{\hat{S}}
\newcommand{\hH}{\hat{H}}
\newcommand{\hbS}{\hat{\bS}}

\newcommand{\bh}{\boldsymbol{h}}

\newcommand{\bsigma}{\boldsymbol{\sigma}}

\begin{document}

\title{Counting Edge Modes with the Higher Berry Curvature:\\[0.05cm] A Bulk Topological Order Parameter for Quantum Spin Chains}

\author{Adam J. McRoberts}
\affiliation{International Centre for Theoretical Physics, Strada Costiera 11, 34151, Trieste, Italy}

\author{Joe Crossley}
\affiliation{School of Physics and Astronomy, University of Nottingham, Nottingham, NG7 2RD, United Kingdom}

\author{Chris Hooley}
\affiliation{Centre for Fluid and Complex Systems, Coventry University, Coventry, CV1 5FB, United Kingdom}

\author{Joe H. Winter}
\affiliation{SUPA, School of Physics and Astronomy, University of St Andrews, North Haugh,\\ St Andrews, Fife, KY16 9SS, United Kingdom}
\affiliation{Max Planck Institute for the Physics of Complex Systems, N\"{o}thnitzer Str. 38, 01187 Dresden, Germany}

\date{\today}
\begin{abstract}
\noindent
We show that the higher Berry curvature (HBC) can be used to count the gapless edge modes created by an entanglement cut, and thus defines an integer-valued topological order parameter for quantum spin chains. Given an individual spin-chain Hamiltonian, we construct an extending family by interpolating to a reference product N\'eel state, and show that the integral of the HBC over this extension is equal to the ordinary Berry phase of half of the chain swept out in response to an \textit{infinitesimal} field. It thus counts the spin of the gapless edge modes exposed by the cut, and a change in its integer value signals a phase transition. We illustrate this with several examples: $S=1/2$, $S=1$, and $S=3/2$ spin-Peierls chains, which undergo `singlet flop' transitions between different patterns of dimerisation; the bilinear-biquadratic chain, which clarifies the connection to the strict symmetry-protected topological phases classification; and the staggered $J_1$--$J_2$ chain, which has both nearest-neighbour and third-neighbour patterns of singlets depending on the signs of the interactions.
\end{abstract}
\maketitle

\begin{figure*}
    \centering
    \includegraphics[width=\linewidth]{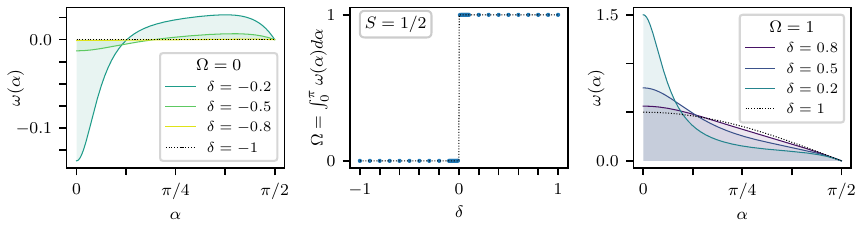}
    \caption{Higher Berry curvature and the integer-valued HBC order parameter for the $S = 1/2$ spin-Peierls chain, with the Hamiltonian given in Eq.~\eqref{eq:spin_Peierls_chain}, where $\delta$ is the staggering between the coupling strengths $J(1 - \delta)$ of odd bonds and $J(1 + \delta)$ of even bonds. (Left) Representative HBC functions $\omega(\alpha)$ for selected values of $\delta$ in the $\Omega = 0$ phase, where a cut through an even bond does not produce any gapless edge modes; and (Right) The same, but for the $\Omega = 1$ phase, where a cut through an even bond produces a protected free spin-$1/2$ edge mode on each side. For a given $\delta$, we use a maximum field strength $h = 1+|\delta|$. (Middle) HBC order parameter $\Omega$, defined by Eqs.~\eqref{eq:extended_family_of_H}-\eqref{eq:HBC_order_parameter_isotropic}, as a function of $\delta$. For $\delta = -0.01$ we obtain $\Omega = 0.00048$, and for $\delta = +0.01$ we obtain $\Omega = 0.99896$; all other obtained values of $\Omega$ are less than $10^{-4}$ away from the correct integer. Each integral~\eqref{eq:HBC_order_parameter_isotropic} is calculated from $401$ sampled values of $\omega(\alpha)$, and the largest Schmidt weight discarded from any state is less than $10^{-8}$.}
    \label{fig:spin_half_Peierls_chain}
\end{figure*}

\section*{Introduction}

Topological phases carry global properties of quantum states that survive local perturbations~\cite{Chenlocal2010, avronHomotopyQuantizationCondensed1983,HasanKane}. This robustness preserves their boundary states, quantised responses~\cite{vonklitzingQuantizedHallEffect1986,konigQuantumSpinHall2007,kohmotoTopologicalInvariantQuantization1985,laughlinQuantizedHallConductivity1981a, thoulessQuantizedHallConductance1982, bernevigQuantumSpinHall2006}, and entanglement structure~\cite{qiGeneralRelationshipEntanglement2012,liEntanglementSpectrumGeneralization2008,pollmannEntanglementSpectrumTopological2010,fidkowskiEntanglementSpectrumTopological2010,alexandradinataTraceIndexSpectral2011}, making such phases promising for fault tolerant quantum information processing and other settings that demand reliable quantum behaviour~\cite{nayakNonAbelianAnyonsTopological2008,kitaevFaulttolerantQuantumComputation2003, kitaevAnyonsExactlySolved2006, kitaevUnpairedMajoranaFermions2001, jeckelmannQuantumHallEffect2003, hartlandQuantumHallEffect1992}.

Topological classifications assign many-body states to algebraic classes~\cite{ryuTopologicalInsulatorsSuperconductors2010, schnyderClassificationTopologicalInsulators2009, kitaevPeriodicTableTopological2009, chiuClassificationTopologicalQuantum2016,chenClassificationGappedSymmetric2011,schuchClassifyingQuantumPhases2011a,fidkowskiTopologicalPhasesFermions2011,ChenClassProper2013}. Their invariants encode each class as a discrete label and compress the local structure, but hide how the invariant is assembled across parameter space. A geometric quantity, however, preserves this local structure and reveals how symmetry constrains the contributions that survive integration. Their integral recovers the global algebraic invariant, allowing geometry to illustrate the relevant classification~\cite{vanderbiltBerryPhasesElectronic2018, bernevigTopologicalInsulatorsTopological2013}.

Topological band theory provides the clearest example of this route from geometry to an algebraic topological index~\cite{ryuTopologicalInsulatorsSuperconductors2010, schnyderClassificationTopologicalInsulators2009, kitaevPeriodicTableTopological2009, chiuClassificationTopologicalQuantum2016}. A Gaussian ground state defines an occupied single-particle subspace whose gauge freedom gives rise to Berry connections, curvatures, and Wilson loops~\cite{marzariMaximallyLocalizedWannier2012,fidkowskiModelCharacterizationGapless2011,zakBerrysPhaseEnergy1989a}. These geometric quantities reveal the obstruction to a globally smooth gauge and show how symmetry constrains its structure across momentum space. Their integrals produce integer invariants such as winding and Chern numbers, while symmetry can identify or quotient these integers to yield reduced classifications such as $\mathbb Z_2$~\cite{Moore2007}.

However, a generic many-body Hamiltonian does not have a single-particle description on which to build this geometric structure~\cite{wangInteractingFermionicTopological2014, verresenOneDimensionalSymmetryProtected2017a, vishwanathPhysicsThreeDimensionalBosonic2013, burnellExactlySolubleModel2014}. In one-dimensional spin chains, the Schmidt states---or equivalently the virtual bond spaces of a matrix product state \cite{schuchClassifyingQuantumPhases2011a}---carry the topological data instead. The projective action of symmetry on this space assigns the ground state its algebraic symmetry-protected-topological (SPT) class~\cite{pollmannSymmetryProtectionTopological2012a, chenClassificationGappedSymmetric2011}. However, it does not supply a local geometric quantity analogous to Berry curvature, and therefore does not show how symmetry shapes the invariant across parameter space. A geometric diagnostic must therefore recover the interacting SPT invariant without assuming its algebraic form in advance.

Higher Berry curvature (HBC) provides this missing geometric structure for interacting ground states~\cite{kapustin2020higher,ohyamaHigherStructuresMatrix2024,ohyamaHigherBerryConnection2025,ohyamaHigherBerryPhase2025,sommerHigherBerryCurvature2025,sommerHigherBerryCurvature2025a,shiozakiHigherBerryCurvature2025}. For a closed three-parameter family of one-dimensional gapped ground states, it defines a local 3-form over parameter space. This local density measures the transfer of ordinary Berry curvature across an entanglement cut~\cite{xuedaBBC2023}, while its integral detects a quantised obstruction to choosing matrix product state data smoothly over the full family~\cite{sommerHigherBerryCurvature2025,ohyamaHigherStructuresMatrix2024}. Existing formulations therefore attach the invariant to the chosen family rather than to any single spin chain.

Here, we use higher Berry curvature to define a diagnostic of an individual spin chain. In \S~\ref{sec:higher_Berry_curvature}, starting from one Hamiltonian, we construct a closed three-parameter extension that has the original Hamiltonian at the poles and connects it, via a staggered field, to trivial Néel product states at the equator. These product states fix the reference phase, so the integrated HBC measures the obstruction to deforming the original ground state into that trivial state. The HBC integral therefore defines a relative bulk invariant of the spin chain itself.

We show in \S~\ref{sec:counting_the_edge_modes} that
the resulting bulk HBC invariant carries a direct boundary meaning: Stokes’ theorem reduces the integral of the HBC over $S^3$ to the ordinary Berry curvature over $S^2$ at its poles. This measures the Berry phase of the spin chain in response to an infinitesimal field, and thus directly counts the gapless edge modes exposed by the cut. The HBC integral recovers this edge response entirely from bulk data for the chosen cut and reference state, and its parity recovers the SPT invariant.
 
In \S~\ref{sec:spin-Peierls_chain} we test this correspondence in bond-alternating $S=1/2$, $S=1$, and $S=3/2$ spin chains~\cite{pytte1974peierls,cross1979new}, where the valence-bond picture makes the boundary spin explicit. The HBC invariant resolves several integer values as the number of singlets across the cut changes. We then clarify the connection between this HBC order parameter and the SPT invariant in \S~\ref{sec:spt_classification}. We show, with the example of the bilinear-biquadratic chain, that any change in the HBC order parameter---even if the SPT invariant does not change---indicates a phase transition.

Lastly, in \S~\ref{sec:J1-J2_chain} we turn to the frustrated $J_1$--$J_2$ chain, where the relevant valence-bond structure is sometimes rather weak, and not always manifest in the Hamiltonian. Nevertheless, the HBC invariant distinguishes the gapped phases through their boundary spin content. Together, all these results establish higher Berry curvature as a geometric route from bulk many-body data to the algebraic index and topological boundary content of quantum spin chains.

\section{Higher Berry curvature \label{sec:higher_Berry_curvature}}

\subsection{Definition of the HBC}

Let us first give a formal definition of the higher Berry curvature in quantum spin chains. Let $\mathcal{M}$ be a compact 3-manifold; this is the external parameter space. Let $\left\{|\Psi(p)\rangle \,:\, p\in\mathcal{M}\right\}$ be a smooth family of normalised states over $\mathcal{M}$. 

Let $L$, $R$ be a partition of the chain into left and right halves. Then for each state we have the Schmidt decomposition, 
\eqa{
|\Psi\rangle = \sum_\eta c_\eta\, |\psi^{(\eta)}_L\rangle |\psi^{(\eta)}_{R}\rangle,
\label{eq:Schmidt_decomposition}
}
and the higher Berry curvature is a 3-form that measures the flow of Berry curvature from $L$ to $R$~\cite{kapustin2020higher,xuedaBBC2023,sommerHigherBerryCurvature2025},
\eqa{
\omega_{L\to R} = -i \sum_\eta d(c_\eta^2) \wedge d\langle\psi^{(\eta)}_L|\wedge d|\psi^{(\eta)}_L\rangle.
\label{eq:HBC_formal_expression}
}
We explain in the Appendix how this can be explicitly calculated when the family of states are expressed as matrix product states (MPS). 

Now, if we further assume that each state in the family is the unique, gapped ground state of some local Hamiltonian (in the MPS language, that the state is \textit{injective}), then the integral of the HBC 3-form over $\mathcal{M}$ is quantised (the Dixmier-Douady number)~\cite{artymowicz2024quantization},
\eqa{
\frac{1}{2\pi}\int_{\mathcal{M}} \omega \,\in\, \mathbb{Z}.
}

More generally, higher Berry curvature can be defined in $d$ spatial dimensions with a $d+2$-dimensional external parameter space~\cite{kapustin2020higher,sommerHigherBerryCurvature2025a}. In this paper, however, we consider only one-dimensional spin chains.

\subsection{Motivating example: dimerised spin-$\mathbf{1/2}$ chains}

Perhaps the simplest example of a family of states with non-trivial higher Berry curvature is the ground states of the following set of Hamiltonians, with the parameter space $\mathcal{M} = S^3$,
\eqa{
\hH_0(\alpha, \theta, \phi) = \;&\Theta\left(\frac{\pi}{2}-\alpha\right)|\cos\alpha| \sum_{n\in\mathbb{Z}} \hbS_{2n}\cdot\hbS_{2n+1} \nn \\
+ \,&\Theta\left(\alpha - \frac{\pi}{2}\right)|\cos\alpha| \sum_{n\in\mathbb{Z}} \hbS_{2n-1}\cdot\hbS_{2n} \nn \\
&+ \sin\alpha\sum_{n\in\mathbb{Z}}\bh(\theta, \phi)\cdot\left(\hbS_{2n} - \hbS_{2n+1}\right),
\label{eq:fully_dimerised_Hamiltonian}
}
where ${\bh(\theta, \phi) = (\sin\theta\cos\phi, \sin\theta\sin\phi, \cos\theta)}$. This model is exactly solvable: for ${\alpha < \pi/2}$, which we will refer to as the northern hemisphere, there are non-zero couplings only on the even bonds ($2n, 2n+1$); and for ${\alpha > \pi/2}$, which we will refer to as the southern hemisphere, there are non-zero couplings only on the odd bonds ($2n-1, 2n$).   

Without loss of generality, we will group together sites on odd bonds into two-site unit cells. Any bipartition of the system is therefore assumed to cut across an even bond (and all such cuts are equivalent due to the two-site/one-cell translation symmetry).

The HBC 3-form of this model over its parameter space is~\cite{sommerHigherBerryCurvature2025}
\eqa{
\omega_0 = \frac{1}{2}\,\Theta\left(\frac{\pi}{2}-\alpha\right) \cos\alpha\,\sin\theta\;d\alpha \wedge d\theta \wedge d\phi,
\label{eq:fully_dimerised_HBC}
}
which, when integrated, returns the non-trivial value
\eqa{
\Omega_0 = \frac{1}{2\pi}\int_{S^3} \omega_0 = 1.
\label{eq:fully_dimerised_HBC_integral}
}

Ref.~\cite{sommerHigherBerryCurvature2025} explicitly showed that this integral is resistant to small deformations of the Hamiltonian family \eqref{eq:fully_dimerised_Hamiltonian}.

\subsection{Topological order parameter from the HBC}

However, by integrating over both hemispheres, we have lost the fact that, intuitively, the two hemispheres---the two different patterns of dimerisation---correspond to two distinct phases. Which phase we call trivial and which non-trivial is, of course, an arbitrary choice (cf. the two phases of the Su-Schrieffer-Heeger model~\cite{su1979solitons,su1980soliton}), but is fixed by our convention that the sites on odd bonds are grouped together to form unit cells---and the cut, therefore, always crosses an even bond between the unit cells.

Let us start by considering the polar Hamiltonians: $\hH_0^N = \hH_0(\alpha = 0)$ at the north pole, where singlets lie between the unit cells; and $\hH_0^S = H_0(\alpha = \pi)$ at the south pole, where the singlets lie within the unit cells. We expect that $\hH_0^N$ should be in the non-trivial phase, and $\hH_0^S$ in the trivial phase, but how do we construct an invariant that will distinguish them?

For this simple, exactly solvable model, there is an obvious solution: the HBC 3-form \eqref{eq:fully_dimerised_HBC} is non-zero only on the northern hemisphere. Integrating only over the hemisphere that contains the Hamiltonian of interest will give us the two values of the topological order parameter that we seek.

However, for application to less trivial cases, we need to define the order parameter such that this selection of the hemisphere is automatic---that is, without prior knowledge of which phase should be non-trivial. 

We thus define a topological order parameter from the higher Berry curvature in the following way: let $\hH$ be the Hamiltonian of a spin chain with (at least) two-site translation symmetry and a unique gapped ground state. To this Hamiltonian, we then associate the extending family ${\hH \mapsto \hH(\alpha, \theta, \phi; h)}$,
\eqa{
\hH(\alpha, \theta, \phi; h) &= |\cos\alpha|\,\hH + \sin\alpha \sum_n (-1)^n \bh(\theta, \phi) \cdot \hbS_n, \nn \\
\bh(\theta, \phi) &= h\,(\sin\theta\cos\phi, \sin\theta\sin\phi, \cos\theta),
\label{eq:extended_family_of_H}
}
where $\hbS_n = (\hS^x_n, \hS^y_n, \hS^z_n)$ in the appropriate spin-$S$ representation. We will also refer to this as `the extension of $\hH$', or just `the extension'. This family of Hamiltonians, then, defines a family of ground states $|\Psi(\alpha, \theta, \phi)\rangle$. For $\alpha = 0$ or $\pi$, this is the ground state of the original Hamiltonian $\hH$, and on the equator $\alpha = \pi/2$ it is a N\'eel product state.

We assume that this extending family is smooth---that is, for any direction $(\theta, \phi)$, the application of the staggered field does not trigger a phase transition at any finite field strength. There is a large class of antiferromagnetic spin chains for which this holds---including, at least, all those whose long-wavelength physics is described by the $O(3)$ non-linear sigma model~\cite{haldane1983continuum}. 

Let us be clear about the differences between this family and the family of states defined by Eq.~\eqref{eq:fully_dimerised_Hamiltonian}. By construction, the two hemispheres of Eq.~\eqref{eq:extended_family_of_H} are identical---we have the same state, the ground state of $\hH$, at \textit{both} the north and south poles.

This reflects our aim---namely, to probe the properties of the single Hamiltonian $\hH$. The extension~\eqref{eq:extended_family_of_H} is necessary to define the HBC 3-form, but we care about it only to the extent it tells us about $\hH$. This contrasts with previous studies of the HBC, which have often been concerned with the global properties of entire families~\cite{xuedaBBC2023,ohyama2024discrete,ohyamaHigherStructuresMatrix2024,ohyamaHigherBerryConnection2025,ohyamaHigherBerryPhase2025,shiozakiHigherBerryCurvature2025,sommerHigherBerryCurvature2025,sommerHigherBerryCurvature2025a,shiozaki2026equivariant}.

So, the HBC 3-form \eqref{eq:HBC_formal_expression} can then be calculated for each state in the extending family~\eqref{eq:extended_family_of_H}, 
\eqa{
|\Psi(\alpha, \theta, \phi)\rangle \;\;\longrightarrow\;\; \omega(\alpha, \theta, \phi)\,d\alpha \wedge d\theta \wedge d\phi,
}
and its integral over this parameter space,
\eqa{
\Omega = \frac{1}{4\pi} \int_{S^3} \omega(\alpha, \theta, \phi) \,d\alpha \wedge d\theta \wedge d\phi,
\label{eq:HBC_order_parameter}
}
is an integer-valued bulk topological order parameter. We have divided by $4\pi$ because the extending family~\eqref{eq:extended_family_of_H} has two copies of $\hH$. Note that the magnitude $h = |\bh(\theta, \phi)|$ of the applied staggered field affects the HBC 3-form, but will not change the integral.

To connect this back to the exactly solvable example, let ${\hH = \hH_0(\alpha_0, \theta_0, \phi_0)}$ be any of the Hamiltonians in the family given by Eq.~\eqref{eq:fully_dimerised_Hamiltonian} (with \textit{fixed} $\alpha_0$, $\theta_0$, $\phi_0$). For any point in the northern hemisphere $\alpha_0 < \pi/2$ (not just the north pole), the extending family~\eqref{eq:extended_family_of_H} gives us two copies of the northern hemisphere, glued together at the equator (the reference product N\'eel state), and the HBC order parameter defined by Eqs.~\eqref{eq:extended_family_of_H}--\eqref{eq:HBC_order_parameter} is $\Omega^N=1$. And, similarly, for any point in the southern hemisphere $\alpha_0 > \pi/2$, we end up integrating over two copies of the southern hemisphere, and $\Omega^S = 0$.

In the remainder of this paper, we will assume $\hH$ has full $SU(2)$ isotropy. In that case, $\omega(\alpha, \theta, \phi) = \omega(\alpha)\sin\theta$, and Eq.~\eqref{eq:HBC_order_parameter} simplifies to a single integral,
\eqa{
\Omega = 2\int_0^{\pi/2} \omega(\alpha)\,d\alpha.
\label{eq:HBC_order_parameter_isotropic}
}

\section{Counting the edge modes \label{sec:counting_the_edge_modes}}

Before we turn to the numerical computation of this order parameter to classify the phases of interacting spin chains, let us first discuss its physical interpretation: what does it mean to have $\Omega \neq 0$?

Mathematically,
\eqa{
\Omega = \frac{1}{4\pi}\int_{S^3} \omega \neq 0
}
means that $\omega$ is not globally exact; that is, there is no $C^1$-smooth (continuously-differentiable) $2$-form $F$ defined everywhere on $S^3$ such that $\omega = dF$.

Now, formally, the HBC 3-form $\omega$ defined in Eq.~\eqref{eq:HBC_formal_expression} can be obtained as the exterior derivative of the 2-form
\eqa{
F = -i \sum_\eta c_\eta^2 \,d\langle\psi^{(\eta)}_L|\wedge d|\psi^{(\eta)}_L\rangle,
\label{eq:primitive_2_form}
}
which is the ordinary Berry curvature of the left half of the chain. If $\Omega \neq 0$, this identification must fail somewhere. 

In isotropic spin chains, if it fails, it fails at the poles. Indeed, we must have ${F = f(\alpha) \sin\theta\,d\theta \wedge d\phi}$. If $f(0)$ or $f(\pi) \neq 0$, then $F$ is ill-defined at the poles---it retains dependence on $\theta$ and $\phi$, even though they are redundant co-ordinates.

So, let $B_N(\epsilon)$, $B_S(\epsilon)$ denote $\epsilon$-neighbourhoods of the north and south poles of $S^3$, respectively. And denote by ${\mathcal{M}_\epsilon = S^3\backslash(B_N(\epsilon) \cup B_S(\epsilon))}$ the parameter space with these small neighbourhoods of the poles removed. By Stokes' theorem, then, we have
\eqa{
\Omega 
\;&=\; \lim_{\epsilon\to0}\,\frac{1}{4\pi}\int_{\mathcal{M}_{\epsilon}} \!\!\omega 
\;=\; \lim_{\epsilon\to0}\,\frac{1}{4\pi}\int_{\pd\mathcal{M}_{\epsilon}} \!\!F
\nn \\[0.1cm]
\;&=\; \lim_{\epsilon\to0}\,\frac{1}{4\pi}\int_{S^2} \left(F\big|_{\alpha\,=\,\pi-\epsilon} - F\big|_{\alpha\,=\,\epsilon}\right) \nn \\
\;&=\; \lim_{\epsilon\to0}\,\frac{1}{2\pi}\int_{S^2} F\big|_{\alpha\,=\,\pi-\epsilon}
\label{eq:HBC=BP}
}

\begin{figure}
    \centering
    \includegraphics[width=\linewidth]{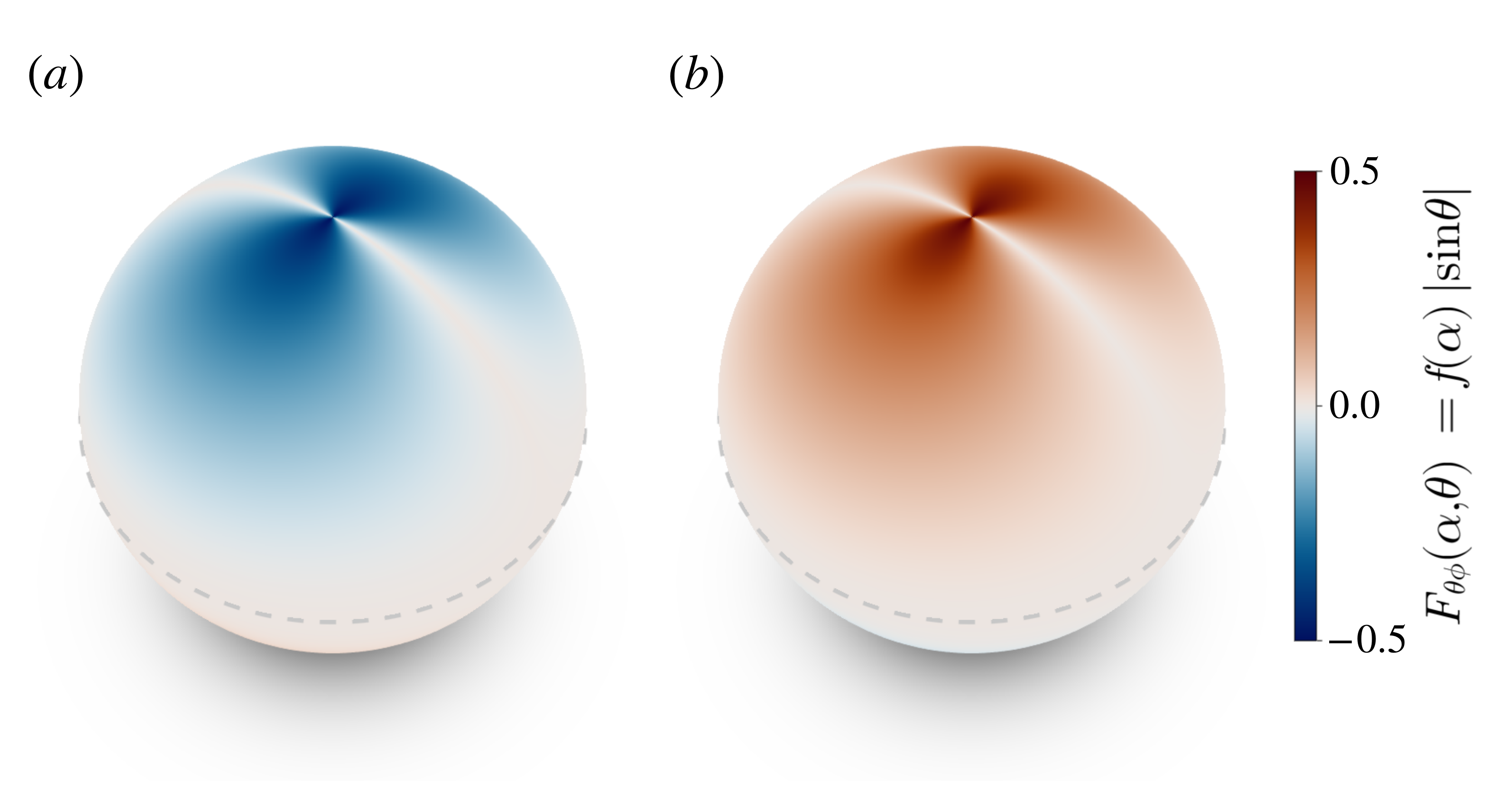}
    \caption{Singularities in $F$, the $2$-form Berry curvature for the left-half of the chain, for the exactly solvable model $\hH = \hH_0^N$. This Hamiltonian has $\Omega = 1$, meaning $F$ (Eq.~\eqref{eq:H_north_pole_F}) is ill-defined at (at least one of) the poles of $S^3$. In the images shown, the polar angle is $\alpha$, and the azimuthal angle is $\theta$ ($\phi$ cannot be shown). The function plotted is $f(\alpha, \theta)$, where $F = f(\alpha, \theta)\,d\theta\wedge d\phi$. A view of the singularity at the north pole $\alpha = 0$ is shown in (a); and the image is rotated by $\pi$ in (b) to give a view of the south pole $\alpha = \pi$. Note that to obtain a closed 2-sphere we have, just for these images, extended the range of $\theta$ from $[0, \pi]$ to $[0, 2\pi)$, and correspondingly cut the range of $\phi$ from $[-\pi, \pi)$ to $[0, \pi]$.}
    \label{fig:F_singularity}
\end{figure}

Thus, the HBC order parameter $\Omega$ is the Berry phase of the left half of the chain swept out in response to an \textit{infinitesimal} applied field. Physically, it counts the gapless edge modes created by cutting the chain, because only the gapless modes can respond to the infinitesimal field.

To make this concrete, consider again the fully-dimerised example, and let $\hH = \hH_0^N$. Then $\Omega = 1$, with $\omega = \frac{1}{2}|\cos\alpha| \sin\theta\,d\alpha \wedge d\theta \wedge d\phi$. The singlets lie between the unit cells, and the cut produces a free spin-$1/2$ on either side:

\begin{figure}[H]
\centering
\begin{tikzpicture}
    \draw[line width=0.4mm] (-0.4, 0) -- (0.0, 0);
    \draw[line width=0.4mm] (0.8, 0) -- (1.6, 0);
    \draw[line width=0.4mm] (2.4, 0) -- (3.2, 0);
    \draw[line width=0.4mm] (4.0, 0) -- (4.8, 0);
    \draw[line width=0.4mm] (5.6, 0) -- (6.0, 0);
    \draw[line width=0.4mm, dotted] (-0.8, 0) -- (-0.5, 0);
    \draw[line width=0.4mm, dotted] (6.1, 0) -- (6.4, 0);
    \draw[fill=green!40] (0, 0) circle (4pt);
    \draw[fill=green!40] (0.8, 0) circle (4pt);
    \draw[fill=green!40] (1.6, 0) circle (4pt);
    \draw[fill=red] (2.4, 0) circle (4pt);
    \draw[fill=red] (3.2, 0) circle (4pt);
    \draw[fill=green!40] (4, 0) circle (4pt);
    \draw[fill=green!40] (4.8, 0) circle (4pt);
    \draw[fill=green!40] (5.6, 0) circle (4pt);
    \draw[decorate,decoration={zigzag, amplitude=0.4mm}, line width=0.4mm] (2.8, -0.6) -- (2.8, 0.8);
    \draw (2.8, -0.7) node[below] {$\Omega = 1$};
\end{tikzpicture}.
\end{figure}

\noindent The Berry curvature $2$-form for the left half of the chain is
\eqa{
F = f(\alpha) \sin\theta\,d\theta \wedge d\phi, \;\;
f(\alpha) = \begin{cases}
    \frac{\sin\alpha - 1}{2} & \alpha < \pi/2 \\
    \frac{1 - \sin\alpha}{2} & \alpha > \pi/2
\end{cases},
\label{eq:H_north_pole_F}
}
which has singularities at the poles, shown in Fig.~\ref{fig:F_singularity}. And, cf. Eq.~\eqref{eq:HBC=BP}, $\Omega$ indeed counts the Berry phase of a single free spin-$1/2$,
\eqa{
\Omega = \underbrace{(f(\pi)-f(0))}_{=1}\frac{1}{4\pi} \int_{S^2} \sin\theta\,d\theta \wedge d\phi = 1.
}
In the opposite case where $\hH = \hH_0^S$, cutting the chain between the unit cells does not produce any edge modes, and $\omega = 0$ (and $F = 0$) everywhere in parameter space:
\begin{figure}[H]
\centering
\begin{tikzpicture}
    \draw[line width=0.4mm] (0.0, 0) -- (0.8, 0);
    \draw[line width=0.4mm] (2.4, 0) -- (1.6, 0);
    \draw[line width=0.4mm] (4.0, 0) -- (3.2, 0);
    \draw[line width=0.4mm] (5.6, 0) -- (4.8, 0);
    
    \draw[line width=0.4mm, dotted] (-0.8, 0) -- (-0.5, 0);
    \draw[line width=0.4mm, dotted] (6.1, 0) -- (6.4, 0);

    \draw[fill=green!40] (0, 0) circle (4pt);
    \draw[fill=green!40] (0.8, 0) circle (4pt);
    \draw[fill=green!40] (1.6, 0) circle (4pt);
    \draw[fill=green!40] (2.4, 0) circle (4pt);
    \draw[fill=green!40] (3.2, 0) circle (4pt);
    \draw[fill=green!40] (4, 0) circle (4pt);
    \draw[fill=green!40] (4.8, 0) circle (4pt);
    \draw[fill=green!40] (5.6, 0) circle (4pt);
    
    \draw[decorate,decoration={zigzag, amplitude=0.4mm}, line width=0.4mm] (2.8, -0.6) -- (2.8, 0.8);
    \draw (2.8, -0.7) node[below] {$\Omega = 0$};
\end{tikzpicture}.
\end{figure}

The utility of the HBC 3-form is that it computes the Berry curvature carried by the entanglement boundary directly from bulk many-body data. For a fixed cut and reference state, the resulting integer resolves the full spin content of the Schmidt boundary states, including contributions that a strict, group-cohomological SPT classification may identify as equivalent. We will return to this last point in \S~\ref{sec:spt_classification}.

\section{Spin-Peierls chain \label{sec:spin-Peierls_chain}}

We now show that the higher Berry curvature order parameter we have defined in Eqs.~\eqref{eq:extended_family_of_H}--\eqref{eq:HBC_order_parameter} can be used to classify the phases of interacting spin chains beyond the exactly solvable example we have discussed so far. 

For the first example, we will consider the spin-Peierls chain~\cite{pytte1974peierls,cross1979new},
\eqa{
\hH = J\sum_n \left(1 + (-1)^n\delta \right)\hbS_n\cdot\hbS_{n+1},
\label{eq:spin_Peierls_chain}
}
for $S = 1/2$, $S = 1$, and $S = 3/2$. We assume $J > 0$ (i.e., that the spins are coupled antiferromagnetically), and, for an array of values of the staggering $\delta \in [-1, 1]$, we compute the HBC order parameter $\Omega(\delta)$. 

For each $\delta$, we select an evenly-spaced grid of values of $\alpha$ on which we will sample the function $\omega(\alpha)$. At each $\alpha$, we begin with a random uMPS at a small bond dimension, and optimise the state at that fixed bond dimension using the VUMPS algorithm~\cite{zauner2018variational} packaged in MPSKit~\cite{Devos_MPSKit_2026}. If the smallest Schmidt weight $c_\eta^2$ is above some threshold (see figure captions), the bond dimension is increased and the state is re-optimised. This process is repeated until at least two Schmidt weights are below the threshold, and the state is finally re-optimised at that bond dimension until the VUMPS residual is below $10^{-10}$.

\begin{figure}
    \centering
    \includegraphics[]{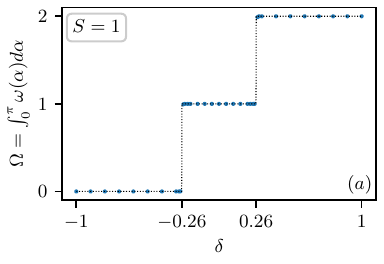}
    \includegraphics[]{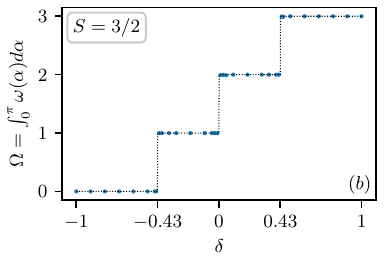}
    \caption{HBC order parameter as a function of the bond-strength staggering $\delta$ for the (a) $S = 1$ and (b) $S = 3/2$ spin-Peierls chains, cf. Eq.~\eqref{eq:spin_Peierls_chain}. As expected, the HBC order parameter measures the spin of the protected gapless edge modes in each phase, cf. the discussion in the main text. The maximum field strength in all cases is $h = 1.0$. Each integral~\eqref{eq:HBC_order_parameter_isotropic} is calculated from $1001$ sampled values of $\omega(\alpha)$, and the largest Schmidt weight discarded from any state is less than $10^{-8}$.}
    \label{fig:spin_one_and_three_half_Peierls_chains}
\end{figure}

The ground state of $\hH(\alpha + \Delta\alpha)$ is then computed at the same bond dimension, starting from the ground state found for $\hH(\alpha)$. In all cases, $\Delta\alpha = 10^{-6}$, and the field direction is, without loss of generality, taken to be $+\bx$. From these two states, the value of the function $\omega(\alpha)$ is calculated using the method detailed in the Appendix. The obtained samples of $\omega(\alpha)$ are interpolated to a continuous function by SciPy's CubicSpline function~\cite{virtanen2020scipy}, which is then integrated to give the value of the HBC order parameter~\eqref{eq:HBC_order_parameter_isotropic}.

\begin{figure}
    \centering
    \includegraphics[]{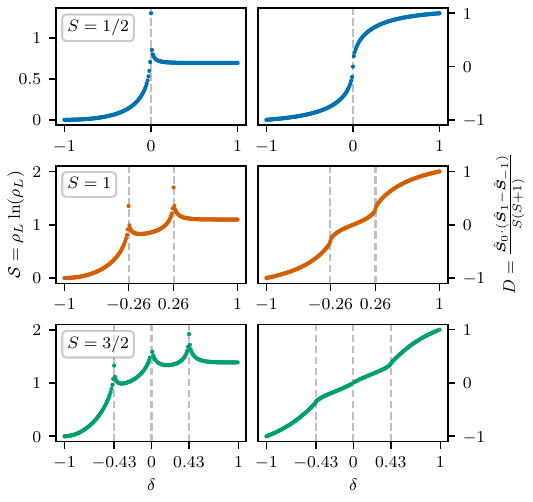}
    \caption{
    Entanglement entropy (left) and dimer-order (right) as a function of bond-strength staggering $\delta$ in the ${S = 1/2}$, $S = 1$, and $S = 3/2$ spin-Peierls chains~\eqref{eq:spin_Peierls_chain}, for comparison with the HBC order parameter (Figs.~\ref{fig:spin_half_Peierls_chain} \&~\ref{fig:spin_one_and_three_half_Peierls_chains}). The spikes in the entanglement entropy at the critical points are clear signatures of the transitions, and their locations are in agreement with those found by the HBC order parameter. The dimer-order does not show the transition points quite as clearly, though it does have features reminiscent of non-analyticities at the correct points. The largest Schmidt weight discarded from any state is less than $10^{-8}$.}
    \label{fig:ent_entropy_dimer_order}
\end{figure}

The results for the HBC order parameter (Figs.~\ref{fig:spin_half_Peierls_chain} \&~\ref{fig:spin_one_and_three_half_Peierls_chains}) are compared with the von Neumann entanglement entropy and dimer-order $D = \langle\hbS_0\cdot(\hbS_1 - \hbS_{-1})\rangle$ in Fig.~\ref{fig:ent_entropy_dimer_order}; all observables agree on the locations of the transitions for all values of $S$.

\subsection{Spin-$\mathbf{1/2}$}

For $S = 1/2$, we have $\hH(\delta = 1) = \hH_0^N$ with $\Omega = 1$; and $\hH(\delta = -1) = \hH_0^S$, with $\Omega = 0$. Then, because it is a topological invariant, we expect the HBC order parameter should not change in the vicinity of these exactly solvable points: the free spin-1/2 edge modes that would be created by cutting the chain in the $\Omega = 1$ phase will no longer be perfectly localised to a single site, but will still exist.

Further, the precise form of $\omega(\alpha)$ will change, but a smooth deformation of the external parameter space can restore it to the form it has at the fixed points $\delta = \pm1$.

The transition between the two phases occurs at the gapless critical point $\delta = 0$, where $\hH$ is the $S = 1/2$ antiferromagnetic Heisenberg chain described by the $\mathfrak{\hat{su}}(2)_1$ CFT, with central charge $c = 1$~\cite{affleck1987critical}. 

The results are shown in Fig.~\ref{fig:spin_half_Peierls_chain}. We find excellent numerical quantisation of the HBC order parameter $\Omega$, even close to the critical point: for $\delta = -0.01$ we obtain $\Omega = 0.00048$, and for $\delta = +0.01$ we obtain $\Omega = 0.99896$; all other values of $\Omega(\delta)$ are less than $10^{-4}$ away from the correct integer.

\subsection{Spin-$\mathbf{1}$}

For the $S=1$ spin-Peierls chain, $\delta = 0$ (which is the $S = 1$ Heisenberg chain) is no longer a gapless point, cf. the Haldane conjecture~\cite{haldane1983continuum,affleck1989quantum}. It is adiabatically connected to the AKLT model~\cite{affleck1987rigorous}, for which the ground state can be expressed as
\begin{figure}[H]
\centering
\begin{tikzpicture}
\centering
    \draw[line width=0.2mm] (-0.6, -0.13) -- (0, -0.13);      
    \draw[line width=0.2mm] (0, 0.13) -- (1.2, 0.13);

    \draw[line width=0.2mm] (1.2, -0.13) -- (2.4, -0.13);      
    \draw[line width=0.2mm] (2.4, 0.13) -- (3.6, 0.13);

    \draw[line width=0.2mm] (3.6, -0.13) -- (4.8, -0.13);      
    \draw[line width=0.2mm] (4.8, 0.13) -- (6, 0.13);
        
    \draw[line width=0.2mm] (6, -0.13) -- (6.6, -0.13);

    \draw[line width=0.4mm, dotted] (-1.1, 0) -- (-0.8, 0);
    \draw[line width=0.4mm, dotted] (6.8, 0) -- (7.1, 0);

    \draw (0, 0) circle (9pt);
    \draw[fill=green!40] (0, -0.13) circle (3pt);
    \draw[fill=green!40] (0, 0.13) circle (3pt);

    \draw (1.2, 0) circle (9pt);
    \draw[fill=green!40] (1.2, -0.13) circle (3pt);
    \draw[fill=green!40] (1.2, 0.13) circle (3pt);

    \draw (2.4, 0) circle (9pt);
    \draw[fill=green!40] (2.4, -0.13) circle (3pt);
    \draw[fill=red] (2.4, 0.13) circle (3pt);

    \draw (3.6, 0) circle (9pt);
    \draw[fill=green!40] (3.6, -0.13) circle (3pt);
    \draw[fill=red] (3.6, 0.13) circle (3pt);

    \draw (4.8, 0) circle (9pt);
    \draw[fill=green!40] (4.8, -0.13) circle (3pt);
    \draw[fill=green!40] (4.8, 0.13) circle (3pt);

    \draw (6, 0) circle (9pt);
    \draw[fill=green!40] (6, -0.13) circle (3pt);
    \draw[fill=green!40] (6, 0.13) circle (3pt);

    \draw[decorate,decoration={zigzag, amplitude=0.4mm}, line width=0.4mm] (3, -0.6) -- (3, 0.8);
    \draw (3, -0.7) node[below] {$\Omega = 1$};
\end{tikzpicture},
\end{figure}

\noindent where each small filled circle represents a virtual spin-$1/2$ degree of freedom, and the larger circles represent the projector onto the $S = 1$ subspace (such that we recover the original Hilbert space). Each spin-$1$ site has one virtual spin-$1/2$ singlet bond with each of its neighbours, and cutting through it produces a spin-$1/2$ edge mode on each side.

At the end-points $\delta = \pm 1$, the model is again fully-dimerised; at $\delta = -1$, the ground state is singlets lying within the unit cells:
\begin{figure}[H]
\centering
\begin{tikzpicture}
\centering
    \draw[line width=0.2mm] (-0.6, -0.13) -- (0, -0.13);      
    \draw[line width=0.2mm] (-0.6, 0.13) -- (0, 0.13);

    \draw[line width=0.2mm] (1.2, -0.13) -- (2.4, -0.13);      
    \draw[line width=0.2mm] (1.2, 0.13) -- (2.4, 0.13);

    \draw[line width=0.2mm] (3.6, -0.13) -- (4.8, -0.13);      
    \draw[line width=0.2mm] (3.6, 0.13) -- (4.8, 0.13);
        
    \draw[line width=0.2mm] (6, -0.13) -- (6.6, -0.13);
    \draw[line width=0.2mm] (6, 0.13) -- (6.6, 0.13);

    \draw[line width=0.4mm, dotted] (-1.1, 0) -- (-0.8, 0);
    \draw[line width=0.4mm, dotted] (6.8, 0) -- (7.1, 0);

    \draw (0, 0) circle (9pt);
    \draw[fill=green!40] (0, -0.13) circle (3pt);
    \draw[fill=green!40] (0, 0.13) circle (3pt);

    \draw (1.2, 0) circle (9pt);
    \draw[fill=green!40] (1.2, -0.13) circle (3pt);
    \draw[fill=green!40] (1.2, 0.13) circle (3pt);

    \draw (2.4, 0) circle (9pt);
    \draw[fill=green!40] (2.4, -0.13) circle (3pt);
    \draw[fill=green!40] (2.4, 0.13) circle (3pt);

    \draw (3.6, 0) circle (9pt);
    \draw[fill=green!40] (3.6, -0.13) circle (3pt);
    \draw[fill=green!40] (3.6, 0.13) circle (3pt);

    \draw (4.8, 0) circle (9pt);
    \draw[fill=green!40] (4.8, -0.13) circle (3pt);
    \draw[fill=green!40] (4.8, 0.13) circle (3pt);

    \draw (6, 0) circle (9pt);
    \draw[fill=green!40] (6, -0.13) circle (3pt);
    \draw[fill=green!40] (6, 0.13) circle (3pt);

    \draw[decorate,decoration={zigzag, amplitude=0.4mm}, line width=0.4mm] (3, -0.6) -- (3, 0.8);
    \draw (3, -0.7) node[below] {$\Omega = 0$};
\end{tikzpicture}
\end{figure}

\noindent And at $\delta = +1$, the ground state is singlets lying between the unit cells. Cutting through them produces a free spin-$1$ mode,
\begin{figure}[H]
\centering
\begin{tikzpicture}
\centering
    \draw[line width=0.2mm] (0, -0.13) -- (1.2, -0.13);      
    \draw[line width=0.2mm] (0, 0.13) -- (1.2, 0.13);

    \draw[line width=0.2mm] (2.4, -0.13) -- (3.6, -0.13);      
    \draw[line width=0.2mm] (2.4, 0.13) -- (3.6, 0.13);

    \draw[line width=0.2mm] (4.8, -0.13) -- (6, -0.13);      
    \draw[line width=0.2mm] (4.8, 0.13) -- (6, 0.13);

    \draw[line width=0.4mm, dotted] (-1.1, 0) -- (-0.8, 0);
    \draw[line width=0.4mm, dotted] (6.8, 0) -- (7.1, 0);

    \draw (0, 0) circle (9pt);
    \draw[fill=green!40] (0, -0.13) circle (3pt);
    \draw[fill=green!40] (0, 0.13) circle (3pt);

    \draw (1.2, 0) circle (9pt);
    \draw[fill=green!40] (1.2, -0.13) circle (3pt);
    \draw[fill=green!40] (1.2, 0.13) circle (3pt);

    \draw (2.4, 0) circle (9pt);
    \draw[fill=red] (2.4, -0.13) circle (3pt);
    \draw[fill=red] (2.4, 0.13) circle (3pt);

    \draw (3.6, 0) circle (9pt);
    \draw[fill=red] (3.6, -0.13) circle (3pt);
    \draw[fill=red] (3.6, 0.13) circle (3pt);

    \draw (4.8, 0) circle (9pt);
    \draw[fill=green!40] (4.8, -0.13) circle (3pt);
    \draw[fill=green!40] (4.8, 0.13) circle (3pt);

    \draw (6, 0) circle (9pt);
    \draw[fill=green!40] (6, -0.13) circle (3pt);
    \draw[fill=green!40] (6, 0.13) circle (3pt);

    \draw[decorate,decoration={zigzag, amplitude=0.4mm}, line width=0.4mm] (3, -0.6) -- (3, 0.8);
    \draw (3, -0.7) node[below] {$\Omega = 2$};
\end{tikzpicture}.
\end{figure}

The transitions between these phases are, again, critical points described by the $\mathfrak{\hat{su}}(2)_1$ CFT, and occur at ${\delta^* \approx \pm 0.26}$~\cite{kitazawa1996phase}. 

The obtained values of $\Omega(\delta)$ are shown in Fig.~\ref{fig:spin_one_and_three_half_Peierls_chains}(a). Again, we have excellent quantisation, even close to the critical points: for the first we have ${\Omega(-0.27) = 0.0021}$ and ${\Omega(-0.25) = 0.9949}$; and for the second, we have ${\Omega(+0.25) = 1.0007}$ and ${\Omega(+0.27) = 1.9898}$. All other values are within $10^{-3}$ of the correct integer.

\subsection{Spin-$\mathbf{3/2}$}

For $S = 3/2$, it is again easiest to visualise the possible phases using an AKLT-like picture of three virtual spin-$1/2$ sites projected onto their $S = 3/2$ subspace. Then, in the fully-dimerised limit $\delta = -1$, we have:
\begin{figure}[H]
\centering
\begin{tikzpicture}
    \draw[line width=0.2mm] (-0.6, -0.2) -- (0, -0.2);  
    \draw[line width=0.2mm] (-0.6, 0) -- (0, 0);    
    \draw[line width=0.2mm] (-0.6, 0.2) -- (0, 0.2);

    \draw[line width=0.2mm] (1.2, -0.2) -- (2.4, -0.2);  
    \draw[line width=0.2mm] (1.2, 0) -- (2.4, 0);    
    \draw[line width=0.2mm] (1.2, 0.2) -- (2.4, 0.2);

    \draw[line width=0.2mm] (3.6, -0.2) -- (4.8, -0.2);  
    \draw[line width=0.2mm] (3.6, 0) -- (4.8, 0);    
    \draw[line width=0.2mm] (3.6, 0.2) -- (4.8, 0.2);
        
    \draw[line width=0.2mm] (6, -0.2) -- (6.6, -0.2);
    \draw[line width=0.2mm] (6, 0) -- (6.6, 0);
    \draw[line width=0.2mm] (6, 0.2) -- (6.6, 0.2);

    \draw[line width=0.4mm, dotted] (-1.1, 0) -- (-0.8, 0);
    \draw[line width=0.4mm, dotted] (6.8, 0) -- (7.1, 0);

    \draw (0, 0) circle (10pt);
    \draw[fill=green!40] (0, -0.2) circle (2.2pt);
    \draw[fill=green!40] (0, 0) circle (2.2pt);
    \draw[fill=green!40] (0, 0.2) circle (2.2pt);

    \draw (1.2, 0) circle (10pt);
    \draw[fill=green!40] (1.2, -0.2) circle (2.2pt);
    \draw[fill=green!40] (1.2, 0) circle (2.2pt);
    \draw[fill=green!40] (1.2, 0.2) circle (2.2pt);

    \draw (2.4, 0) circle (10pt);
    \draw[fill=green!40] (2.4, -0.2) circle (2.2pt);
    \draw[fill=green!40] (2.4, 0) circle (2.2pt);
    \draw[fill=green!40] (2.4, 0.2) circle (2.2pt);

    \draw (3.6, 0) circle (10pt);
    \draw[fill=green!40] (3.6, -0.2) circle (2.2pt);
    \draw[fill=green!40] (3.6, 0) circle (2.2pt);
    \draw[fill=green!40] (3.6, 0.2) circle (2.2pt);

    \draw (4.8, 0) circle (10pt);
    \draw[fill=green!40] (4.8, -0.2) circle (2.2pt);
    \draw[fill=green!40] (4.8, 0) circle (2.2pt);
    \draw[fill=green!40] (4.8, 0.2) circle (2.2pt);

    \draw (6, 0) circle (10pt);
    \draw[fill=green!40] (6, -0.2) circle (2.4pt);
    \draw[fill=green!40] (6, 0) circle (2.4pt);
    \draw[fill=green!40] (6, 0.2) circle (2.4pt);

    \draw[decorate,decoration={zigzag, amplitude=0.4mm}, line width=0.4mm] (3, -0.6) -- (3, 0.8);
    \draw (3, -0.7) node[below] {$\Omega = 0$};
\end{tikzpicture},
\end{figure}

\noindent and in the opposite limit $\delta = +1$:
\begin{figure}[H]
\centering
\begin{tikzpicture}
    \draw[line width=0.2mm] (0, -0.2) -- (1.2, -0.2);  
    \draw[line width=0.2mm] (0, 0) -- (1.2, 0);    
    \draw[line width=0.2mm] (0, 0.2) -- (1.2, 0.2);

    \draw[line width=0.2mm] (2.4, -0.2) -- (3.6, -0.2);  
    \draw[line width=0.2mm] (2.4, 0) -- (3.6, 0);    
    \draw[line width=0.2mm] (2.4, 0.2) -- (3.6, 0.2);

    \draw[line width=0.2mm] (4.8, -0.2) -- (6, -0.2);  
    \draw[line width=0.2mm] (4.8, 0) -- (6, 0);    
    \draw[line width=0.2mm] (4.8, 0.2) -- (6, 0.2);

    \draw[line width=0.4mm, dotted] (-1.1, 0) -- (-0.8, 0);
    \draw[line width=0.4mm, dotted] (6.8, 0) -- (7.1, 0);

    \draw (0, 0) circle (10pt);
    \draw[fill=green!40] (0, -0.2) circle (2.2pt);
    \draw[fill=green!40] (0, 0) circle (2.2pt);
    \draw[fill=green!40] (0, 0.2) circle (2.2pt);

    \draw (1.2, 0) circle (10pt);
    \draw[fill=green!40] (1.2, -0.2) circle (2.2pt);
    \draw[fill=green!40] (1.2, 0) circle (2.2pt);
    \draw[fill=green!40] (1.2, 0.2) circle (2.2pt);

    \draw (2.4, 0) circle (10pt);
    \draw[fill=red] (2.4, -0.2) circle (2.2pt);
    \draw[fill=red] (2.4, 0) circle (2.2pt);
    \draw[fill=red] (2.4, 0.2) circle (2.2pt);

    \draw (3.6, 0) circle (10pt);
    \draw[fill=red] (3.6, -0.2) circle (2.2pt);
    \draw[fill=red] (3.6, 0) circle (2.2pt);
    \draw[fill=red] (3.6, 0.2) circle (2.2pt);

    \draw (4.8, 0) circle (10pt);
    \draw[fill=green!40] (4.8, -0.2) circle (2.2pt);
    \draw[fill=green!40] (4.8, 0) circle (2.2pt);
    \draw[fill=green!40] (4.8, 0.2) circle (2.2pt);

    \draw (6, 0) circle (10pt);
    \draw[fill=green!40] (6, -0.2) circle (2.4pt);
    \draw[fill=green!40] (6, 0) circle (2.4pt);
    \draw[fill=green!40] (6, 0.2) circle (2.4pt);

    \draw[decorate,decoration={zigzag, amplitude=0.4mm}, line width=0.4mm] (3, -0.6) -- (3, 0.8);
    \draw (3, -0.7) node[below] {$\Omega = 3$};
\end{tikzpicture}.
\end{figure}

Then, by analogy with the $\Omega = 1$ phase of the $S = 1$ case, we expect that the $S = 3/2$ spin-Peierls chain will realise two partially-dimerised phases:
\begin{figure}[H]
\centering
\begin{tikzpicture}
    \draw[line width=0.2mm] (-0.6, -0.2) -- (0, -0.2);  
    \draw[line width=0.2mm] (-0.6, 0) -- (0, 0);    
    \draw[line width=0.2mm] (0, 0.2) -- (1.2, 0.2);

    \draw[line width=0.2mm] (1.2, -0.2) -- (2.4, -0.2);  
    \draw[line width=0.2mm] (1.2, 0) -- (2.4, 0);    
    \draw[line width=0.2mm] (2.4, 0.2) -- (3.6, 0.2);

    \draw[line width=0.2mm] (3.6, -0.2) -- (4.8, -0.2);  
    \draw[line width=0.2mm] (3.6, 0) -- (4.8, 0);    
    \draw[line width=0.2mm] (4.8, 0.2) -- (6, 0.2);
        
    \draw[line width=0.2mm] (6, -0.2) -- (6.6, -0.2);
    \draw[line width=0.2mm] (6, 0) -- (6.6, 0);

    \draw[line width=0.4mm, dotted] (-1.1, 0) -- (-0.8, 0);
    \draw[line width=0.4mm, dotted] (6.8, 0) -- (7.1, 0);

    \draw (0, 0) circle (10pt);
    \draw[fill=green!40] (0, -0.2) circle (2.2pt);
    \draw[fill=green!40] (0, 0) circle (2.2pt);
    \draw[fill=green!40] (0, 0.2) circle (2.2pt);

    \draw (1.2, 0) circle (10pt);
    \draw[fill=green!40] (1.2, -0.2) circle (2.2pt);
    \draw[fill=green!40] (1.2, 0) circle (2.2pt);
    \draw[fill=green!40] (1.2, 0.2) circle (2.2pt);

    \draw (2.4, 0) circle (10pt);
    \draw[fill=green!40] (2.4, -0.2) circle (2.2pt);
    \draw[fill=green!40] (2.4, 0) circle (2.2pt);
    \draw[fill=red] (2.4, 0.2) circle (2.2pt);

    \draw (3.6, 0) circle (10pt);
    \draw[fill=green!40] (3.6, -0.2) circle (2.2pt);
    \draw[fill=green!40] (3.6, 0) circle (2.2pt);
    \draw[fill=red] (3.6, 0.2) circle (2.2pt);

    \draw (4.8, 0) circle (10pt);
    \draw[fill=green!40] (4.8, -0.2) circle (2.2pt);
    \draw[fill=green!40] (4.8, 0) circle (2.2pt);
    \draw[fill=green!40] (4.8, 0.2) circle (2.2pt);

    \draw (6, 0) circle (10pt);
    \draw[fill=green!40] (6, -0.2) circle (2.4pt);
    \draw[fill=green!40] (6, 0) circle (2.4pt);
    \draw[fill=green!40] (6, 0.2) circle (2.4pt);

    \draw[decorate,decoration={zigzag, amplitude=0.4mm}, line width=0.4mm] (3, -0.6) -- (3, 0.8);
    \draw (3, -0.7) node[below] {$\Omega = 1$};
\end{tikzpicture},
\end{figure}

\noindent and

\begin{figure}[H]
\centering
\begin{tikzpicture}
    \draw[line width=0.2mm] (-0.6, -0.2) -- (0, -0.2);  
    \draw[line width=0.2mm] (0, 0) -- (1.2, 0);    
    \draw[line width=0.2mm] (0, 0.2) -- (1.2, 0.2);

    \draw[line width=0.2mm] (1.2, -0.2) -- (2.4, -0.2);  
    \draw[line width=0.2mm] (2.4, 0) -- (3.6, 0);    
    \draw[line width=0.2mm] (2.4, 0.2) -- (3.6, 0.2);

    \draw[line width=0.2mm] (3.6, -0.2) -- (4.8, -0.2);  
    \draw[line width=0.2mm] (4.8, 0) -- (6, 0);    
    \draw[line width=0.2mm] (4.8, 0.2) -- (6, 0.2);
        
    \draw[line width=0.2mm] (6, -0.2) -- (6.6, -0.2);

    \draw[line width=0.4mm, dotted] (-1.1, 0) -- (-0.8, 0);
    \draw[line width=0.4mm, dotted] (6.8, 0) -- (7.1, 0);

    \draw (0, 0) circle (10pt);
    \draw[fill=green!40] (0, -0.2) circle (2.2pt);
    \draw[fill=green!40] (0, 0) circle (2.2pt);
    \draw[fill=green!40] (0, 0.2) circle (2.2pt);

    \draw (1.2, 0) circle (10pt);
    \draw[fill=green!40] (1.2, -0.2) circle (2.2pt);
    \draw[fill=green!40] (1.2, 0) circle (2.2pt);
    \draw[fill=green!40] (1.2, 0.2) circle (2.2pt);

    \draw (2.4, 0) circle (10pt);
    \draw[fill=green!40] (2.4, -0.2) circle (2.2pt);
    \draw[fill=red] (2.4, 0) circle (2.2pt);
    \draw[fill=red] (2.4, 0.2) circle (2.2pt);

    \draw (3.6, 0) circle (10pt);
    \draw[fill=green!40] (3.6, -0.2) circle (2.2pt);
    \draw[fill=red] (3.6, 0) circle (2.2pt);
    \draw[fill=red] (3.6, 0.2) circle (2.2pt);

    \draw (4.8, 0) circle (10pt);
    \draw[fill=green!40] (4.8, -0.2) circle (2.2pt);
    \draw[fill=green!40] (4.8, 0) circle (2.2pt);
    \draw[fill=green!40] (4.8, 0.2) circle (2.2pt);

    \draw (6, 0) circle (10pt);
    \draw[fill=green!40] (6, -0.2) circle (2.4pt);
    \draw[fill=green!40] (6, 0) circle (2.4pt);
    \draw[fill=green!40] (6, 0.2) circle (2.4pt);

    \draw[decorate,decoration={zigzag, amplitude=0.4mm}, line width=0.4mm] (3, -0.6) -- (3, 0.8);
    \draw (3, -0.7) node[below] {$\Omega = 2$};
\end{tikzpicture}.
\end{figure}

The transition between $\Omega = 1$ and $\Omega = 2$ occurs at $\delta = 0$, where the model is the $S = 3/2$ Heisenberg chain. Again, cf. Haldane's conjecture, this is critical and described by $\mathfrak{\hat{su}}(2)_1$. The other transitions occur at $\delta^* \approx \pm0.43$~\cite{kitazawa1997phase}, and are also critical points described by $\mathfrak{\hat{su}}(2)_1$.

The obtained values of $\Omega(\delta)$ are shown in Fig.~\ref{fig:spin_one_and_three_half_Peierls_chains}(b). The quantisation in this case is not quite as robust as at smaller values of $S$, owing to increasing numerical costs with the enlarged on-site Hilbert space, but is still very clear---all points, even those closest to the transitions, are less than $0.01$ away from the correct integer.

\section{SPT classification and the HBC \label{sec:spt_classification}}

We now return to the question we deferred at the end of \S~\ref{sec:counting_the_edge_modes}: what is the relation between the HBC order parameter and the strict, group-cohomological classification of SPT phases?

The spin chains we are considering have the following discrete symmetries: time-reversal (${\hbS \mapsto -\hbS}$), and spatial-inversion about the entanglement cut. The SPT phases of this $\mathbb{Z}_2^T \times \mathbb{Z}_2^I$ discrete symmetry group are classified by a $\mathbb{Z}_2$-valued SPT index $\mathcal{K} = \pm1$. Essentially, it measures whether the entanglement cut exposes a half-integer-spin edge mode ($\mathcal{K} = -1$) or an integer-spin edge mode ($\mathcal{K} = +1$)~\cite{chenClassificationGappedSymmetric2011,schuchClassifyingQuantumPhases2011a,pollmannSymmetryProtectionTopological2012a}.

The HBC order parameter, by contrast, takes values in $\mathbb{Z}$, and, as shown in \S~\ref{sec:counting_the_edge_modes}, counts the spin of the edge modes by measuring the Berry phase response to an infinitesimal staggered field. Clearly, then, the two are related by
\eqa{
\mathcal{K} = (-1)^{\Omega}.
}

However, this raises the question: if $\Omega$ jumps by an even number, such that $\mathcal{K}$ is unchanged, is there necessarily a phase transition? To take a particular example, the $\Omega = 0$ and $\Omega = 2$ phases of the $S = 1$ spin-Peierls chain (Fig.~\ref{fig:spin_one_and_three_half_Peierls_chains}) are, under the SPT classification, identified as the same SPT-trivial phase $\mathcal{K} = +1$.

\begin{figure}
    \centering
    \includegraphics[]{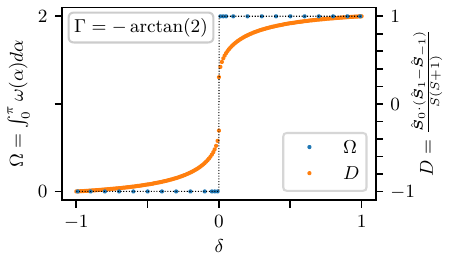}
    \caption{HBC and dimer order parameters for the bilinear-biquadratic chain \eqref{eq:bilinear_biquadratic_chain} at $\Gamma = -\arctan(2) \approx -1.107$. There is a first-order transition at $\delta = 0$, where there is spontaneous dimerisation, and the dimer order parameter is discontinuous. The HBC order parameter $\Omega$ jumps from $0$ to $2$ across this transition. Each value of $\Omega$ was obtained from $1001$ sampled values of $\omega(\alpha)$, and the largest Schmidt weight discarded from any state was less than $10^{-8}$. As discussed in the main text, a strict SPT classification would consider the two sides to be same SPT phase---but, if only transformations for which the HBC is well-defined are permitted, any change in $\Omega$ always requires a phase transition.}
    \label{fig:blbq_chain}
\end{figure}

The answer lies in the assumption we have used to define the HBC order parameter $\Omega(\hH)$, which we re-iterate here for emphasis: the extension \eqref{eq:extended_family_of_H} of $\hH$ defines a \textit{smooth} family of gapped ground states. In other words, for any direction $(\theta, \phi)$ of staggered field, $\hH$ has no phase transitions as a function of field strength (i.e., as a function of $\alpha$), except possibly at $\alpha = 0$. 

This protecting assumption is what enhances the classification from $\mathbb{Z}_2$ to $\mathbb{Z}$. Each $\hH(\delta)$ carries its own extension \eqref{eq:extended_family_of_H}, and for the HBC order parameter to change as $\delta$ is varied, there must be at least one point, somewhere in the extensions, where the HBC becomes ill-defined---a phase transition. But, by the protecting assumption, this must happen at $\alpha = 0$; that is, the phase transition must occur in the base Hamiltonians $\hH(\delta)$.

Indeed, one standard way to show that the $\Omega = 0$ and $\Omega = 2$ cases are the same SPT-trivial $\mathcal{K} = +1$ phase is to add single-ion anisotropy,
\eqa{
\hH(\delta, D_z) = J\sum_n\left(1 + (-1)^n\delta \right)\hbS_n\cdot\hbS_{n+1} + D_z \sum_n (\hS_n^z)^2.
}
If we take $D_z/J \gtrsim 0.968$, even $\delta = 0$ is no longer in the Haldane phase, and we encounter no phase transitions as we sweep $\delta$ from $-1$ to $1$~\cite{tonegawa1996ground}. But the single-ion anisotropy violates the protecting assumption: there are now phase transitions at finite applied staggered field. 

Put another way, given the assumption of smoothness, the HBC order parameter can, via Stokes' theorem, be meaningfully considered a property of a single Hamiltonian---under these circumstances, it gives an enhanced classification of the topological phases. Without the protecting assumption, the HBC order parameter is inextricably a property of the entire extended family, and the integer value classifies obstructions to transforming entire families into each other---there is then no reason to expect it to be equivalent to the the SPT classification of individual Hamiltonians.

An example of a phase transition where $\Omega$ jumps by an even number is given by the $S = 1$ bilinear-biquadratic chain,
\eqa{
\hH = \sum_n \,(1 + (-1)^n\delta)\, \Bigl(\!&\cos\Gamma\, \hbS_n\cdot\hbS_{n+1} \nn \\[-0.2cm] &+ \sin\Gamma\, (\hbS_n\cdot\hbS_{n+1})^2 \Bigr).
\label{eq:bilinear_biquadratic_chain}
}
For $\delta = 0$ and $-\pi/4 < \Gamma < \pi/4$, this model is in the Haldane phase ($\mathcal{K} = -1$, $\Omega = 1$). For ${-3\pi/4 < \Gamma < -\pi/4}$, the model is spontaneously dimerised~\cite{schollwock1996onset}: a small negative $\delta$ selects the $\Omega = 0$ phase, and a small positive $\delta$ selects the $\Omega = 2$ phase; there is a first-order transition at $\delta = 0$. We show the HBC order parameter and dimer order as a function of $\delta$ for $\Gamma = -\arctan(2) \approx-1.107$ in Fig.~\ref{fig:blbq_chain}.

\section{Staggered $\mathbf{J_1}$--$\mathbf{J_2}$ chain \label{sec:J1-J2_chain}}

As a final example, we now consider the spin-$1/2$ staggered $J_1$--$J_2$ chain, with the Hamiltonian
\eqa{
\hH = J_1 \sum_n \left(1 + (-1)^n\delta \right)\hbS_n\cdot\hbS_{n+1} + J_2 \sum_n \hbS_n\cdot\hbS_{n+2}.
}
We will consider only the case where the second-neighbour interactions are antiferromagnetic, $J_2 > 0$.

\subsection{$\mathbf{J_1 > 0}$}

Let us first consider the case where both couplings are antiferromagnetic, $J_1 > 0$. For $\delta = 0$ (no staggering), there is spontaneous dimerisation associated with the formation of nearest-neighbour singlets~\cite{haldane1982spontaneous,white1996dimerization}, which persists until the transition into a critical antiferromagnetic phase at $J_1/J_2 \approx 4.15$~\cite{okamoto1992fluid,nomura1994critical}. At the Majumdar-Ghosh point $J_1/J_2 = 2$, the two nearest-neighbour singlet covers are exact degenerate ground states~\cite{majumdar1969next,majumdar1969nextii}.

A non-zero value of $\delta$ simply selects one of these patterns of singlets; so, for $J_1 > 0$ we have the same topological phases as the $S = 1/2$ Peierls chain, and obtain $\Omega = 0, 1$ for $\delta < 0$ and $\delta > 0$, respectively.

\subsection{$\mathbf{J_1 < 0}$}

For ferromagnetic nearest-neighbour couplings, $J_1 < 0$, there is also spontaneous dimerisation at $\delta = 0$; but in this case it is associated with the formation of nearest-neighbour spin-$1$ \textit{triplets}~\cite{furukawa2012ground}, persisting until the transition to a ferromagnetic phase at $J_1/J_2 = -4$~\cite{sirker2011j}. 

Again, a non-zero value of $\delta$ selects one of these patterns of triplet dimerisation, so, overall, we expect four phases in the region $|J_1|/J_2 \leq 1$, $-1 \leq \delta \leq 1$.

\begin{figure}
    \centering
    \includegraphics[width=\linewidth]{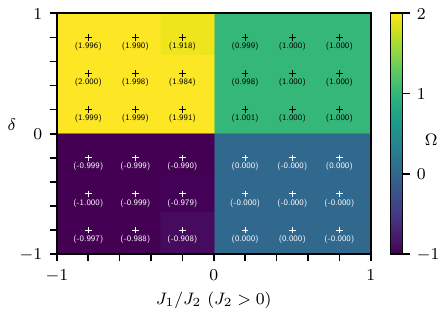}
    \includegraphics[width=\linewidth]{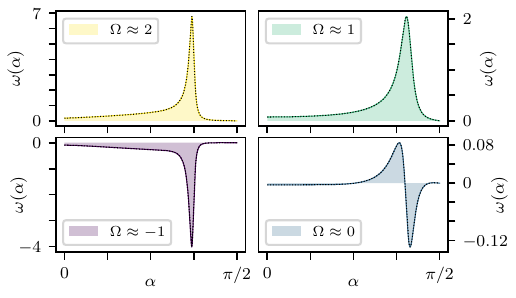}
    \caption{(Upper) Phase diagram of the staggered $J_1$--$J_2$ spin chain measured by the HBC order parameter; the + markers denote the points where the HBC was numerically computed, and the obtained value (to three decimal places) is written below. The colour plot is then schematically filled by interpolating to the nearest data point. The vertical axis ${J_1 = 0}$ (for which $\delta$ is redundant) corresponds to two decoupled antiferromagnetic Heisenberg chains. The horizontal axis ${\delta = 0}$ is a line of first-order transitions. (Lower) Plots of the HBC function $\omega(\alpha)$ for $J_1/J_2 = \pm0.5$, $\delta = \pm 0.5$. For $J_1 > 0$, each HBC integral is calculated with $401$ values of $\alpha$, and the largest discarded Schmidt weight is less than $10^{-7}$. For $J_1 < 0$, we use $1001$ values of $\alpha$, and a Schmidt weight tolerance of $10^{-8}$.}
    \label{fig:J1-J2_phase_diagram}
\end{figure}

Let us now try to reason out the values of the HBC order parameter for the phases where $J_1 < 0$. After the formation of the nearest-neighbour triplets, the state is a complicated superposition of the different ways to pair up spins in neighbouring triplets, such that every triplet has one bond with each neighbouring triplet (as in the standard AKLT construction).

One might expect the weight of the singlets to be highest on the second-neighbour $J_2$ bonds, which would lead to further spontaneous dimerisation---but this is not the favoured pattern: Ref.~\cite{agrapidis2019coexistence} provides evidence that \textit{order-by-disorder} instead selects, albeit very weakly, a \textit{third}-neighbour pattern of singlet formation. 

That is, the ground state for $\delta < 0$ is adiabatically-connected to the $\mathcal{D}_3$-singlet state,

\begin{figure}[H]
\centering
\begin{tikzpicture}
    \draw[dotted, line width=0.3mm] (0, 0.0) -- (0.6, 0.6);
    \draw[dotted, line width=0.3mm] (1.2, 0.0) -- (1.8, 0.6);
    \draw[dotted, line width=0.3mm] (2.4, 0.0) -- (3.0, 0.6);
    \draw[dotted, line width=0.3mm] (3.6, 0.0) -- (4.2, 0.6);
    \draw[dotted, line width=0.3mm] (4.8, 0.0) -- (5.4, 0.6);
    \draw[dotted, line width=0.3mm] (6.0, 0.0) -- (6.6, 0.6);

    \draw[line width=0.5mm] (0, 0.0) -- (1.8, 0.6);
    \draw[line width=0.5mm] (1.2, 0.0) -- (3.0, 0.6);
    \draw[line width=0.5mm] (2.4, 0.0) -- (4.2, 0.6);
    \draw[line width=0.5mm] (3.6, 0.0) -- (5.4, 0.6);
    \draw[line width=0.5mm] (4.8, 0.0) -- (6.6, 0.6);
    \draw[line width=0.5mm] (-0.3, 0.3) -- (0.6, 0.6);
    \draw[line width=0.5mm] (6.0, 0.0) -- (6.9, 0.3);

    \draw[line width=0.4mm, dotted] (-0.8, 0.3) -- (-0.5, 0.3);
    \draw[line width=0.4mm, dotted] (7.1, 0.3) -- (7.4, 0.3);
    
    \draw[fill=green!40] (0, 0) circle (3.5pt);
    \draw[fill=green!40] (0.6, 0.6) circle (3.5pt);
    \draw[fill=green!40] (1.2, 0) circle (3.5pt);
    \draw[fill=green!40] (1.8, 0.6) circle (3.5pt);
    \draw[fill=red] (2.4, 0) circle (3.5pt);
    \draw[fill=green!40] (3.0, 0.6) circle (3.5pt);

    \draw[fill=green!40] (3.6, 0) circle (3.5pt);
    \draw[fill=red] (4.2, 0.6) circle (3.5pt);
    \draw[fill=green!40] (4.8, 0) circle (3.5pt);
    \draw[fill=green!40] (5.4, 0.6) circle (3.5pt);
    \draw[fill=green!40] (6.0, 0) circle (3.5pt);
    \draw[fill=green!40] (6.6, 0.6) circle (3.5pt);

    \draw (5.8, -0.7) node[below] {$J_1 < 0$, $\delta < 0$,};
    \draw[decorate,decoration={zigzag, amplitude=0.4mm}, line width=0.4mm] (3.3, -0.6) -- (3.3, 1.3);
    \draw (3.3, -0.7) node[below] {$\Omega = -1$};
\end{tikzpicture}
\end{figure}
\noindent where the dotted lines denote the bonds on which the triplets form, and the solid lines are the $\mD_3$-singlet pattern.

Note that the sign of the HBC order parameter is now negative, because the free spin in the left half has appeared on an \textit{odd}-numbered site (recall our convention that the cut always goes through an even bond), and the magnetic field is staggered (cf. Eq.~\eqref{eq:extended_family_of_H}).   

For $\delta > 0$, we have the complementary pattern, and the ground state is adiabatically-connected to the $\mD'_3$-singlet state,

\begin{figure}[H]
\centering
\begin{tikzpicture}
    \draw[dotted, line width=0.3mm] (-0.3, 0.3) -- (0.0, 0.0);
    \draw[dotted, line width=0.3mm] (0.6, 0.6) -- (1.2, 0.0);
    \draw[dotted, line width=0.3mm] (2.4, 0.0) -- (1.8, 0.6);
    \draw[dotted, line width=0.3mm] (3.6, 0.0) -- (3.0, 0.6);
    \draw[dotted, line width=0.3mm] (4.8, 0.0) -- (4.2, 0.6);
    \draw[dotted, line width=0.3mm] (6.0, 0.0) -- (5.4, 0.6);
    \draw[dotted, line width=0.3mm] (6.9, 0.3) -- (6.6, 0.6);

    \draw[line width=0.5mm] (-0.3, 0.48) -- (1.2, 0.0);
    \draw[line width=0.5mm] (0.6, 0.6) -- (2.4, 0.0);
    \draw[line width=0.5mm] (1.8, 0.6) -- (3.6, 0.0);
    \draw[line width=0.5mm] (3.0, 0.6) -- (4.8, 0.0);
    \draw[line width=0.5mm] (4.2, 0.6) -- (6.0, 0.0);
    \draw[line width=0.5mm] (5.4, 0.6) -- (6.9, 0.12);

    \draw[line width=0.4mm, dotted] (-0.8, 0.3) -- (-0.5, 0.3);
    \draw[line width=0.4mm, dotted] (7.1, 0.3) -- (7.4, 0.3);
    
    \draw[fill=green!40] (0, 0) circle (3.5pt);
    \draw[fill=green!40] (0.6, 0.6) circle (3.5pt);
    \draw[fill=green!40] (1.2, 0) circle (3.5pt);
    \draw[fill=red] (1.8, 0.6) circle (3.5pt);
    \draw[fill=green!40] (2.4, 0) circle (3.5pt);
    \draw[fill=red] (3.0, 0.6) circle (3.5pt);

    \draw[fill=red] (3.6, 0) circle (3.5pt);
    \draw[fill=green!40] (4.2, 0.6) circle (3.5pt);
    \draw[fill=red] (4.8, 0) circle (3.5pt);
    \draw[fill=green!40] (5.4, 0.6) circle (3.5pt);
    \draw[fill=green!40] (6.0, 0) circle (3.5pt);
    \draw[fill=green!40] (6.6, 0.6) circle (3.5pt);

    \draw (5.8, -0.7) node[below] {$J_1 < 0$, $\delta > 0$.};
    \draw[decorate,decoration={zigzag, amplitude=0.4mm}, line width=0.4mm] (3.3, -0.6) -- (3.3, 1.3);
    \draw (3.3, -0.7) node[below] {$\Omega = 2$};
\end{tikzpicture}
\end{figure}

\noindent The cut produces two free spins in the left half, both on even-numbered sites, and thus we obtain $\Omega = +2$.

The phase diagram of the staggered $J_1$--$J_2$ chain is shown in Fig.~\ref{fig:J1-J2_phase_diagram}, along with plots of $\omega(\alpha)$ for a representative point in each of the four phases. Each phase is characterised by a unique integer value of the HBC order parameter, $-1$, $0$, $1$, or $2$.

\section*{Conclusions}

In this work, we have used the higher Berry curvature to construct a bulk order parameter for quantum spin chains, which directly measures the spin of the gapless edge modes exposed by an entanglement cut. 

For a given spin chain, we have introduced an extending family over which the bulk HBC is integrated---by Stokes' theorem, however, it is equal to the ordinary Berry phase swept out in response to an infinitesimal field, and is therefore meaningfully a property of an individual Hamiltonian.

The spin-Peierls chains provide controlled tests of this correspondence, since their valence-bond limits make the edge spin explicit. Across the $S=1/2$, $S=1$, and $S=3/2$ chains, $\Omega$ follows each change in the number of spin-$1/2$ degrees of freedom exposed by the cut and remains sharply quantised even close to the critical points. The staggered $J_1$--$J_2$ chain is a stronger test, because the relevant singlet structure for $J_1 < 0$ is both (i) rather weak and (ii) not manifest in the microscopic Hamiltonian: nevertheless, the HBC resolves all four gapped sectors, $\Omega=-1,0,1,2$, and distinguishes both the number of edge modes and their positions on the lattice.

We have clarified the correspondence between the integer-valued HBC order parameter and the strict, $\mathbb{Z}_2$-valued group-cohomological SPT invariant---the latter is the parity of the former. The enhanced, integer-valued classification offered by the HBC order parameter reflects a more restrictive set of allowed transformations---those that leave the ground state adiabatically-connected to the reference N\'eel state---and any change in the value of $\Omega$, even if the SPT invariant does not change, requires a phase transition. We have shown an example of this in the bilinear-biquadratic chain.

These results suggest that the construction extends beyond the spin chains studied here. In other chains and spin ladders, an appropriate extension could diagnose phases distinguished by more complex patterns of entanglement. More broadly, the infinitesimal spin field used here may represent one instance of a general strategy in which a symmetry-controlled probe sweeps the quantum numbers carried by an entanglement boundary, allowing higher Berry curvature to recover them from the bulk. Analogous constructions for fermionic, bosonic, and higher-dimensional systems could therefore reveal boundary structure that strict SPT classifications alone do not retain. The present invariant may be the one-dimensional spin realisation of a broader bulk geometry of interacting entanglement boundaries.

Higher Berry curvature thus restores, for interacting spin chains, the route from local geometry to global topology. Its local density reveals how symmetry organises Berry curvature over parameter space, its integral assembles those contributions into a quantised bulk invariant, and its pole curvature identifies the boundary degrees of freedom that make the invariant physical. Higher Berry curvature therefore reads the topology of an interacting spin chain from the bulk, without discarding the boundary spin that gives the phase its physical content.

\begin{acknowledgements}
J.C. was funded by UK Research and Innovation (UKRI) under the UK government’s Horizon Europe funding guarantee [grant number EP/Y036069/1].
\end{acknowledgements}

\subsection*{Appendix: Calculation of the HBC from MPS}

In this appendix, we explain how to calculate the HBC 3-form for a particular state $|\Psi(A)\rangle$, represented as a uniform MPS (uMPS) with tensor $A$. That is, formally,
\eqa{
|\Psi(&A)\rangle = \sum_{\{\sigma_n\}}\left(\prod_n A^{\sigma_n}\right) \left(\bigotimes_n |\sigma_n\rangle\right) \nn \\[0.2cm]
&=\;
\begin{tikzpicture}[
    baseline={(current bounding box.center)},
    tensor/.style={
        draw,
        rectangle,
        minimum width=0.7cm,
        minimum height=0.6cm
    }
]
\node[tensor] (A1) at (0,0) {$A$};
\node[tensor] (A2) at (1.5,0) {$A$};
\node[tensor] (A3) at (3,0) {$A$};
\draw[dotted, thick] (-1.2, 0) -- (-0.7, 0);
\draw (-0.7,0) -- (A1.west);
\draw (A1.east) -- (A2.west);
\draw (A2.east) -- (A3.west);
\draw (A3.east) -- (3.7,0);
\draw[dotted, thick] (3.7, 0) -- (4.2, 0);
\draw (A1.south) -- ++(0,-0.33);
\draw (A2.south) -- ++(0,-0.33);
\draw (A3.south) -- ++(0,-0.33);
\end{tikzpicture},
}
where each $A^{\sigma}$ is a $\chi \times \chi$ matrix, for some bond-dimension $\chi$, and the index $\sigma = 1, ..., d$, where $d$ is the dimension of the local Hilbert space ($d = (2S+1)^2$ for a unit cell of two spin-$S$ sites).  

The below transformation of the formal expression given in Eq.~\eqref{eq:HBC_formal_expression} into explicit matrix operations will largely follow Ref.~\cite{sommerHigherBerryCurvature2025}. However: (i) we explicitly point out the presence of a boundary term omitted therein, and (ii) we explain how symmetry can be exploited to greatly improve the precision of the tangent states.


We aim to calculate the higher Berry curvature at a point $p \in S^3$ with co-ordinates $\alpha$, $\theta$, $\phi$, for the extending family of Hamiltonians given in Eq.~\eqref{eq:extended_family_of_H}. 

We first calculate the ground state of $\hH(\alpha, \theta, \phi)$ and $\hH(\alpha + \Delta\alpha, \theta, \phi)$ using the VUMPS algorithm~\cite{zauner2018variational} as provided by \mbox{MPSKit}~\cite{Devos_MPSKit_2026}. We use an internal convergence error (VUMPS resiudal) of $10^{-10}$, and a finite difference $\Delta\alpha = 10^{-6}$.

We then use the gauge freedom that, for any matrix $M \in \mathrm{GL}(\chi, \mathbb{C})$, the tensor $A^\sigma \mapsto M^{-1}A^{\sigma}M$ represents the same physical state to express the ground state tensors $A$ in left-canonical form. That is, the left-environment is the identity,
\eqa{
{A^{\sigma}}^{\dagger}A^{\sigma} 
\;=\; 
\begin{tikzpicture}[
    baseline={(0, -0.6)},
    tensor/.style={
        draw,
        rectangle,
        minimum width=0.7cm,
        minimum height=0.6cm
    }
]
\node[tensor] (A1) at (0,0) {$A$};
\node[tensor] (A2) at (0,-1) {$A$};
\draw (-0.7,0) -- (A1.west);
\draw (-0.7,-1) -- (A2.west);
\draw (-0.7,-1) -- (-0.7,0);
\draw (A1.south) -- (A2.north);
\draw (A1.east) -- (0.7, 0);
\draw (A2.east) -- (0.7, -1);
\end{tikzpicture}
\;=\;
\begin{tikzpicture}[
    baseline={(0, -0.6)},
    tensor/.style={
        draw,
        rectangle,
        minimum width=0.7cm,
        minimum height=0.6cm
    }
]
\draw (-0.5, 0) -- (0, 0);
\draw (-0.5, -1) -- (0, -1);
\draw (-0.5, -1) -- (-0.5, 0);
\end{tikzpicture}
\;=\;
\mathbb{1},
}
and the right-environment is a diagonal matrix,
\eqa{
A^{\sigma} R\,{A^{\sigma}}^{\dagger}
\;=\;
\begin{tikzpicture}[
    baseline={(0, -0.6)},
    tensor/.style={
        draw,
        rectangle,
        minimum width=0.7cm,
        minimum height=0.6cm
    }
]
\node[tensor] (A1) at (0,0) {$A$};
\node[tensor] (A2) at (0,-1) {$A$};
\node[tensor] (R) at (1,-0.5) {$R$};
\draw (-0.6,0) -- (A1.west);
\draw (-0.6,-1) -- (A2.west);
\draw (A1.south) -- (A2.north);
\draw (A1.east) -- (1, 0);
\draw (A2.east) -- (1, -1);
\draw (R.north) -- (1, 0);
\draw (R.south) -- (1, -1);
\end{tikzpicture}
\;=\;
\begin{tikzpicture}[
    baseline={(0, -0.6)},
    tensor/.style={
        draw,
        rectangle,
        minimum width=0.7cm,
        minimum height=0.6cm
    }
]
\node[tensor] (R) at (1,-0.5) {$R$};
\draw (R.north) -- (1, 0);
\draw (R.south) -- (1, -1);
\draw (0.5, 0) -- (1, 0);
\draw (0.5, -1) -- (1, -1);
\end{tikzpicture}
\;=\; 
R.
}
The entries of the diagonal right-environment $R$ are the squared Schmidt coefficients $c_\eta^2$, cf. Eq.~\eqref{eq:Schmidt_decomposition}. 

Note that, because the original Hamiltonian $\hH$ is assumed to be isotropic, the Schmidt coefficients cannot depend on $\theta$ or $\phi$; the exterior derivative of the right-environment is thus 
\eqa{
dR \,=\, \pd_\alpha R\,d\alpha \,\approx\, \frac{R(\alpha + \Delta\alpha) - R(\alpha)}{\Delta\alpha} \,d\alpha
}

We now consider the exterior derivatives of the individual Schmidt states, $d|\psi_L^{(\eta)}\rangle$. Each state can be represented as a uMPS with a boundary at the right,
\eqa{
|\psi_L^{(\eta)}\rangle 
\;=\;
\begin{tikzpicture}[
    baseline={(current bounding box.center)},
    tensor/.style={
        draw,
        rectangle,
        minimum width=0.7cm,
        minimum height=0.6cm
    }
]
\node[tensor] (A1) at (0,0) {$A$};
\node[tensor] (A2) at (1.5,0) {$A$};
\node[tensor] (A3) at (3,0) {$A$};
\draw[dotted, thick] (-1.2, 0) -- (-0.7, 0);
\draw (-0.7,0) -- (A1.west);
\draw (A1.east) -- (A2.west);
\draw (A2.east) -- (A3.west);
\draw (A3.east) -- (3.8,0);
\draw (3.8, -0.05) node[right] {$\eta$};
\draw (A1.south) -- ++(0,-0.33);
\draw (A2.south) -- ++(0,-0.33);
\draw (A3.south) -- ++(0,-0.33);
\end{tikzpicture}.
}
Then, using the product rule, the exterior derivative of this state can be written as a uMPS tangent vector~\cite{vanderstraeten2019tangent},
\eqa{
d|&\psi_L^{(\eta)}\rangle = \sum_{\{\bsigma\}}\,\sum_m \,\Bigl(\prod_{n<m} \!\!A^{\sigma_n}\Bigr) dA^{\sigma_m} \Bigl(\prod_{n>m}\!\! A^{\sigma_n}\Bigr) |\bsigma\rangle
\nn \\[0.2cm]
&\;=\;\;\;\;
\begin{tikzpicture}[
    baseline={(current bounding box.center)},
    tensor/.style={
        draw,
        rectangle,
        minimum width=0.7cm,
        minimum height=0.6cm
    }
]
\node[tensor] (A1) at (0,0) {$A$};
\node[tensor] (A2) at (1.5,0) {$A$};
\node[tensor] (A3) at (3,0) {$dA$};
\draw[dotted, thick] (-1.2, 0) -- (-0.7, 0);
\draw (-0.7,0) -- (A1.west);
\draw (A1.east) -- (A2.west);
\draw (A2.east) -- (A3.west);
\draw (A3.east) -- (3.8,0);
\draw (3.8, -0.05) node[right] {$\eta$};
\draw (A1.south) -- ++(0,-0.33);
\draw (A2.south) -- ++(0,-0.33);
\draw (A3.south) -- ++(0,-0.33);
\end{tikzpicture}
\nn \\[0.2cm]
&\;\;\;\;\!+\;\;
\begin{tikzpicture}[
    baseline={(current bounding box.center)},
    tensor/.style={
        draw,
        rectangle,
        minimum width=0.7cm,
        minimum height=0.6cm
    }
]
\node[tensor] (A1) at (0,0) {$A$};
\node[tensor] (A2) at (1.5,0) {$dA$};
\node[tensor] (A3) at (3,0) {$A$};
\draw[dotted, thick] (-1.2, 0) -- (-0.7, 0);
\draw (-0.7,0) -- (A1.west);
\draw (A1.east) -- (A2.west);
\draw (A2.east) -- (A3.west);
\draw (A3.east) -- (3.8,0);
\draw (3.8, -0.05) node[right] {$\eta$};
\draw (A1.south) -- ++(0,-0.33);
\draw (A2.south) -- ++(0,-0.33);
\draw (A3.south) -- ++(0,-0.33);
\end{tikzpicture}
\nn \\[0.1cm]
&\;\;\;\;\!+ \;\;\;\; ...\;,
\label{eq:Schmidt_state_tangent_vector}
}
where $dA$ is the exterior derivative of the uMPS tensor. 

Now, we can use the isotropy to construct the tangent vectors exactly, without resorting to calculating two more ground states and taking finite differences. Without loss of generality, we can fix $\bh = h \bx$. Generically, one may expand $dA = \pd_\alpha A\,d\alpha + \pd_\theta A\,d\theta + \pd_\phi A\,d\phi$. We do not need to consider $\pd_\alpha A$ owing to the antisymmetry of the wedge product and the fact that the Schmidt coefficients depend only on $\alpha$; and $\pd_\theta A$ and $\pd_\phi A$ can be obtained to machine precision by applying infinitesimal rotation matrices around the $y$- and $z$-axes, respectively, to the physical index of $A$.

At this point we could start substituting the uMPS expressions into Eq.~\eqref{eq:HBC_formal_expression}; but we can simplify the calculation by making use of the gauge freedom of the uMPS tangent space---with a caveat. The tangent space of a uniform MPS has the gauge freedom~\cite{vanderstraeten2019tangent}
\eqa{
dA^{\sigma} \;\mapsto\; dA^{\sigma} + XA^{\sigma} - A^{\sigma}X,
\label{eq:tangent_space_gauge_freedom}
}
for any matrix $X$. Cf. Eq.~\eqref{eq:Schmidt_state_tangent_vector}, the term $-A^{\sigma}X$ with the exterior derivative acting on site $m$ is cancelled by the term $+XA^{\sigma}$ with it acting on site $m + 1$. The utility of this is that we may transform to the so-called left-tangent space gauge, where
\eqa{
\begin{tikzpicture}[
    baseline={(0, -0.6)},
    tensor/.style={
        draw,
        rectangle,
        minimum width=0.7cm,
        minimum height=0.6cm
    }
]
\node[tensor] (A1) at (0,0) {$dA$};
\node[tensor] (A2) at (0,-1) {$A$};
\draw (-0.7,0) -- (A1.west);
\draw (-0.7,-1) -- (A2.west);
\draw (-0.7,-1) -- (-0.7,0);
\draw (A1.south) -- (A2.north);
\draw (A1.east) -- (0.7, 0);
\draw (A2.east) -- (0.7, -1);
\end{tikzpicture}
\;\;=\;\;
\begin{tikzpicture}[
    baseline={(0, -0.6)},
    tensor/.style={
        draw,
        rectangle,
        minimum width=0.7cm,
        minimum height=0.6cm
    }
]
\node[tensor] (A1) at (0,0) {$A$};
\node[tensor] (A2) at (0,-1) {$dA$};
\draw (-0.7,0) -- (A1.west);
\draw (-0.7,-1) -- (A2.west);
\draw (-0.7,-1) -- (-0.7,0);
\draw (A1.south) -- (A2.north);
\draw (A1.east) -- (0.7, 0);
\draw (A2.east) -- (0.7, -1);
\end{tikzpicture}
\;\;=\;\;
0,
}
which, when we take the inner product $d\langle\psi^{(\eta)}_L|\wedge d|\psi^{(\eta)}_L\rangle$, removes all terms where the exterior derivatives are acting on different sites.

However, Eq.~\eqref{eq:tangent_space_gauge_freedom} is not a gauge freedom of the Schmidt state tangent space due to the boundary; the final $-A^{\sigma}X$ term is not cancelled, and must be added back on.

So, consider $dA_\theta = \pd_\theta A\,d\theta$. Let $X_\theta$ be the matrix that puts this in left-tangent space gauge, cf. Eq.~\eqref{eq:tangent_space_gauge_freedom}, and denote this transformed tangent vector by $\mathcal{B}_\theta = B_\theta\,d\theta$, i.e., $\mathcal{B}_\theta = dA_\theta + X_\theta A - A X_\theta$. Let $B_\phi$ and $X_\phi$ be defined analogously. Then the higher Berry curvature is
\eqa{
i\omega(\alpha) 
\;&=\;
\begin{tikzpicture}[
    baseline={(0, -0.6)},
    tensor/.style={
        draw,
        rectangle,
        minimum width=0.7cm,
        minimum height=0.6cm
    }
]
\node[tensor] (A1) at (0,0) {$B_\theta$};
\node[tensor] (A2) at (0,-1) {$B_\phi$};
\node[tensor] (R) at (3.2, -0.5) {$\pd_\alpha R$};
\node[draw,
    rectangle,
    minimum width=1.0cm,
    minimum height=1.6cm
] (Y) at (1.6, -0.5) {$(\mathbb{1}\!-\!\mathbb{E}_A^A)^{-1}$};
\draw (-0.6,0) -- (A1.west);
\draw (-0.6,-1) -- (A2.west);
\draw (-0.6,-1) -- (-0.6,0);
\draw (A1.south) -- (A2.north);
\draw (A1.east) -- (0.68, 0);
\draw (A2.east) -- (0.68, -1);
\draw (R.north) -- (3.2, 0);
\draw (R.south) -- (3.2, -1);
\draw (3.2, 0) -- (2.52, 0);
\draw (3.2, -1) -- (2.52, -1);
\end{tikzpicture}
\;+\;
\begin{tikzpicture}[
    baseline={(0, -0.6)},
    tensor/.style={
        draw,
        rectangle,
        minimum width=0.7cm,
        minimum height=0.6cm
    }
]
\node[tensor] (A1) at (0,0) {$X_\theta$};
\node[tensor] (A2) at (0,-1) {$X_\phi$};
\node[tensor] (R) at (1,-0.5) {$\pd_\alpha R$};
\draw (-0.6,0) -- (A1.west);
\draw (-0.6,-1) -- (A2.west);
\draw (-0.6,-1) -- (-0.6,0);
\draw (A1.east) -- (1, 0);
\draw (A2.east) -- (1, -1);
\draw (R.north) -- (1, 0);
\draw (R.south) -- (1, -1);
\end{tikzpicture}
\nn \\[0.3cm]
\;&\;-\;
\begin{tikzpicture}[
    baseline={(0, -0.6)},
    tensor/.style={
        draw,
        rectangle,
        minimum width=0.7cm,
        minimum height=0.6cm
    }
]
\node[tensor] (A1) at (0,0) {$B_\phi$};
\node[tensor] (A2) at (0,-1) {$B_\theta$};
\node[tensor] (R) at (3.2, -0.5) {$\pd_\alpha R$};
\node[draw,
    rectangle,
    minimum width=1.0cm,
    minimum height=1.6cm
] (Y) at (1.6, -0.5) {$(\mathbb{1}\!-\!\mathbb{E}_A^A)^{-1}$};
\draw (-0.6,0) -- (A1.west);
\draw (-0.6,-1) -- (A2.west);
\draw (-0.6,-1) -- (-0.6,0);
\draw (A1.south) -- (A2.north);
\draw (A1.east) -- (0.68, 0);
\draw (A2.east) -- (0.68, -1);
\draw (R.north) -- (3.2, 0);
\draw (R.south) -- (3.2, -1);
\draw (3.2, 0) -- (2.52, 0);
\draw (3.2, -1) -- (2.52, -1);
\end{tikzpicture}
\;-\;
\begin{tikzpicture}[
    baseline={(0, -0.6)},
    tensor/.style={
        draw,
        rectangle,
        minimum width=0.7cm,
        minimum height=0.6cm
    }
]
\node[tensor] (A1) at (0,0) {$X_\phi$};
\node[tensor] (A2) at (0,-1) {$X_\theta$};
\node[tensor] (R) at (1,-0.5) {$\pd_\alpha R$};
\draw (-0.6,0) -- (A1.west);
\draw (-0.6,-1) -- (A2.west);
\draw (-0.6,-1) -- (-0.6,0);
\draw (A1.east) -- (1, 0);
\draw (A2.east) -- (1, -1);
\draw (R.north) -- (1, 0);
\draw (R.south) -- (1, -1);
\end{tikzpicture}\;,
}
where $(\mathbb{1} - \mathbb{E}_A^A)^{-1}$ stands for the geometric series
\eqa{
\begin{tikzpicture}[
    baseline={(0, -0.6)},]
\node[draw,
    rectangle,
    minimum width=1.0cm,
    minimum height=1.6cm
] (Y) at (1.6, -0.5) {$(\mathbb{1}\!-\!\mathbb{E}_A^A)^{-1}$};
\draw (0.38, 0) -- (0.68, 0);
\draw (0.38, -1) -- (0.68, -1);
\draw (2.82, 0) -- (2.52, 0);
\draw (2.82, -1) -- (2.52, -1);
\end{tikzpicture}
\;&=\;
\begin{tikzpicture}[
    baseline={(0, -0.6)},]
\draw (-0.3,0) -- (0.3, 0);
\draw (-0.3,-1) -- (0.3, -1);
\end{tikzpicture}
\;+\;
\begin{tikzpicture}[
    baseline={(0, -0.6)},
    tensor/.style={
        draw,
        rectangle,
        minimum width=0.7cm,
        minimum height=0.6cm
    }
]
\node[tensor] (A1) at (0,0) {$A$};
\node[tensor] (A2) at (0,-1) {$A$};
\draw (-0.6,0) -- (A1.west);
\draw (-0.6,-1) -- (A2.west);
\draw (A1.south) -- (A2.north);
\draw (A1.east) -- (0.6, 0);
\draw (A2.east) -- (0.6, -1);
\end{tikzpicture}
\;+\;
\begin{tikzpicture}[
    baseline={(0, -0.6)},
    tensor/.style={
        draw,
        rectangle,
        minimum width=0.7cm,
        minimum height=0.6cm
    }
]
\node[tensor] (A1) at (0,0) {$A$};
\node[tensor] (A2) at (0,-1) {$A$};
\node[tensor] (A3) at (1,0) {$A$};
\node[tensor] (A4) at (1,-1) {$A$};
\draw (-0.6,0) -- (A1.west);
\draw (-0.6,-1) -- (A2.west);
\draw (A1.south) -- (A2.north);
\draw (A3.south) -- (A4.north);
\draw (A1.east) -- (A3.west);
\draw (A2.east) -- (A4.west);
\draw (A3.east) -- (1.6, 0);
\draw (A4.east) -- (1.6, -1);
\end{tikzpicture}
\nn \\[0.2cm]
\;&\;\;\;\;\;+\;
\begin{tikzpicture}[
    baseline={(0, -0.6)},
    tensor/.style={
        draw,
        rectangle,
        minimum width=0.7cm,
        minimum height=0.6cm
    }
]
\node[tensor] (A1) at (0,0) {$A$};
\node[tensor] (A2) at (0,-1) {$A$};
\node[tensor] (A3) at (1,0) {$A$};
\node[tensor] (A4) at (1,-1) {$A$};
\node[tensor] (A5) at (2,0) {$A$};
\node[tensor] (A6) at (2,-1) {$A$};
\draw (-0.6,0) -- (A1.west);
\draw (-0.6,-1) -- (A2.west);
\draw (A1.south) -- (A2.north);
\draw (A3.south) -- (A4.north);
\draw (A5.south) -- (A6.north);
\draw (A1.east) -- (A3.west);
\draw (A2.east) -- (A4.west);
\draw (A3.east) -- (A5.west);
\draw (A4.east) -- (A6.west);
\draw (A5.east) -- (2.6, 0);
\draw (A6.east) -- (2.6, -1);
\end{tikzpicture}
\;+\;\;\;...\;,
}
which concludes the conversion of the formal expression~\eqref{eq:HBC_formal_expression} for the higher Berry curvature into an explicit matrix calculation.

\bibliography{refs}

\end{document}